\RequirePackage{fix-cm,amsmath}
\RequirePackage{rotating}

\documentclass[smallextended]{svjour3}     

\usepackage{cite}

\smartqed  
\usepackage{hyperref}
\usepackage{nomencl}
\usepackage{enumitem}
\usepackage{footnote}
\usepackage{graphicx}
\usepackage{mathtools}
\usepackage{bm}
\usepackage{multirow}
\usepackage{rotating}
\usepackage{pdflscape}
\usepackage{array}

\usepackage{wrapfig}
\usepackage{lscape}
\usepackage{epstopdf}
\usepackage{geometry}
\usepackage{siunitx}

\usepackage[flushleft]{threeparttable}
\usepackage[export]{adjustbox}
\usepackage{newtxtext,txfonts}
\usepackage{booktabs}

\journalname{}
\begin{document}

\title{Apparent Liquid Permeability in Mixed-Wet Shale Permeable Media}
%



\author{Dian Fan
\and Amin Ettehadtavakkol
\and Wendong Wang}


\institute{Dian Fan  \at
Department of Chemical Engineering, University College London, UK
\email{d.fan@ucl.ac.uk}\\
\\
\and
Amin Ettehadtavakkol \at
Bob L. Herd Department of Petroleum Engineering, Texas Tech University, Lubbock, TX, USA               
\and
Wendong Wang \at
School of Petroleum Engineering, China University of Petroleum (East China), Qingdao, China
}

\date{Received: date / Accepted: date}
\maketitle

\begin{abstract}
Apparent liquid permeability (ALP) in ultra-confined permeable media is primarily governed by the pore confinement and fluid-rock interactions. A new ALP model is required to predict the interactive effect of the above two on the flow in mixed-wet, heterogeneous nanoporous media. This study derives an ALP model and integrates the compiled results from molecular dynamics (MD) simulations, scanning electron microscopy, atomic force microscopy, and mercury injection capillary pressure. The ALP model assumes viscous forces, capillary forces, and liquid slippage in tortuous, rough pore throats. Predictions of the slippage of water and octane are validated against MD data reported in the literature. In up-scaling the proposed liquid transport model to the representative-elementary-volume scale, we integrate the geological fractals of the shale rock samples including their pore size distribution, pore throat tortuosity, and pore-surface roughness. 
Sensitivity results for the ALP indicate that when the pore size is below 100 nm pore confinement allows oil to slip in both hydrophobic and hydrophilic pores, yet it also restricts the ALP due to the restricted intrinsic permeability. The ALP reduces to the well-established Carman-Kozeny equation for no-slip viscous flow in a bundle of capillaries, which reveals a distinguishable liquid flow behavior in shales versus conventional rocks. Compared to the Klinkenberg equation, the proposed ALP model reveals an important insight into the similarities and differences between liquid versus gas flow in shales. 
\end{abstract}


\section{Introduction}
Flow enhancement of liquids in confined hydrophilic and hydrophobic nanotubes is often observed in experiments where the liquid flow rate is reported to be several orders of magnitude more than that predicted by the classic Hagen-Poiseuille equation \cite{de2002fluid,joseph2008carbon,myers2011slip,podolska2013water,whitby2007fluid}. Molecular dynamics (MD) simulations are often used to understand the fluid structure and the fast transport mechanisms under confinement \cite{hummer2001water,joseph2008carbon,majumder2005nanoscale,striolo2006mechanism}. Physical properties (e.g. viscosity and density) of liquid near the tube wall can be different from the bulk liquid due to liquid-solid interactions, which is found the main cause for the fast transport of both non-wetting and wetting liquids \cite{falk2010molecular,noy2007nanofluidics}. Fast transport of non-wetting liquid is attributed to the hydrogen bonding of the liquid, which results in the recession of liquid from the solid surface \cite{hummer2001water}, the formation of ``a nearly frictionless vapor interface'' between the surface and the bulk phase \cite{majumder2005nanoscale}, or fast ballistic diffusion of liquid \cite{striolo2006mechanism}. Fast transport of wetting liquid is attributed to the presence of excessive dissolved gas at the liquid-solid interface \cite{de2002fluid} or the capability of water migrating from one adjacent adsorption site to another \cite{ho2011liquid}. 

\begin{table}[t]
\caption{Quantitative analytical models of flow enhancement.}
\label{tab:model}
{\renewcommand{\arraystretch}{1.5} \begin{subequations}\label{eq:f_Tol}
\begin{flalign}
\hline 
& \hspace{-3cm} \text{Authors} \hspace{3.5cm} \text{Flow enhancement models} \nonumber \\\hline
& \hspace{-3.1cm} \text{Tolstoi } \text{\cite{blake1990slip}} \hspace{2.8cm}
 f = 1+\frac{4l_{slip}}{r}   \\
&\hspace{0.9cm} l_{slip}=\delta_0\left[e^{\alpha S \left(W_l-W_{ls}\right) / k_BT}-1\right]
\end{flalign}
\end{subequations}

\begin{subequations}\label{eq:f_Tho}
\begin{flalign}
& \hspace{-3.5cm} \text{Thomas \& McGaughey } \text{\cite{thomas2008reassessing}} 
 \hspace{1.3cm} f = \big(1+\frac{4l_{slip}}{r}\big)\frac{\mu_b}{\mu(r)}   \\
&  \hspace{1cm} \mu(r)=\mu_{w} \frac{A_{w}(r)}{A_{t}(r)}+\mu_{b}\left[1-\frac{A_{w}(r)}{A_{t}(r)}\right] \label{eq:visco}\\
&  \hspace{1cm} l_{slip} = l_{silp,\infty}+\frac{C'}{r^3}
\end{flalign}
\end{subequations}

\begin{subequations}\label{eq:f_Myers}
\begin{flalign}
& \hspace{-1cm} \text{Myers } \text{\cite{myers2011slip}} 
\hspace{2.8cm}  f=1+\frac{4 l_{slip}}{r} \approx 1+ \frac{4\delta_w}{r}\frac{\mu_b}{\mu_w}\\
& \hspace{3.1cm}  l_{slip} = \delta_w\left(\frac{\mu_{b}}{\mu_{w}}-1\right)\left[1-\frac{3}{2} \frac{\delta_w}{r}+\left(\frac{\delta_w}{r}\right)^{2}-\frac{1}{4}\left(\frac{\delta_w}{r}\right)^{3}\right]
\end{flalign}
\end{subequations}

\begin{subequations}\label{eq:f_Matt}
\begin{flalign}
&  \hspace{1.8cm}\text{Mattia \& Calabr\`o } \text{\cite{mattia2012explaining}} 
\hspace{1.7cm}  f=\left(\frac{r-\delta_w}{r}\right)^{4}\left(1-\frac{\mu_{b}}{\mu_{w}}\right)+\frac{\mu_{b}}{\mu_{w}}\left(1+\frac{4l_{slip}}{r}\right) \approx \frac{8\mu_bL}{r^2}\frac{D_s}{W_A}   \\
&\hspace{6cm}  l_{slip}  =\frac{2\mu_wL_s}{r}\frac{D_s}{W_A} \label{eq:lslipr} \\\hline \nonumber
& \multicolumn{1}{p{\textwidth}}{Nomenclature: $l_{slip}$ is the slip length. $\delta_0$ is the distance between the centers of the neighboring liquid molecules. $\delta_w$ is the near-wall region thickness. $W_l$ and $W_{ls}$ denote the work of adhesion of the liquid and the liquid-solid, respectively. $S$ and $\alpha_s$ are the surface area and the fraction of the available sites for liquid migration, respectively. $\mu(r)$ is the weighted-average viscosity over the cross-sectional area fraction of the near-wall region (denoted as $A_w$) and the total flow region (denoted as $A_t$), where their viscosity is denoted by $\mu_w$ and $\mu_b$, respectively. $l_{slip,\infty}$ is the slip length of a liquid on a flat surface (without confinement). $C'$ is a fitting parameter. $L_s$ is the length of the nanotube (straight length).}
\end{flalign}
\end{subequations}}
\end{table}

\begin{figure}[t]
	\centering
	\includegraphics[width=0.5\linewidth]{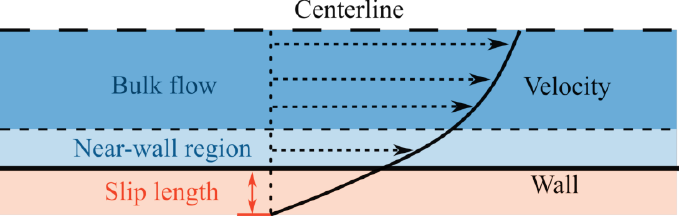}
	\caption{Bulk flow and near-wall regions in a confined channel or pore.}
	\label{fig:Conceptual}
\end{figure}

In principle, MD simulations are the best tool to quantify microscopic physics, yet their computational effort can be intensive and time-consuming. Quantitative analytical models have so far been able to predict the flow enhancement of the confined liquid. The flow enhancement factor ($f$), defined as the ratio of the measured (apparent) volumetric flow rate ($Q_{app}$) to the intrinsic volumetric flow rate predicted by the Hagen-Poiseuille equation ($Q$), is usually applied to evaluate flow enhancement through nanotubes. Table \ref{tab:model} summarizes some classical analytical models for flow enhancement. The main differences between these models are how viscosity is modeled and who is the contributor to the flow enhancement. The Tolstoi model \cite{blake1990slip}, one of the earliest quantitative attempted to model liquid slippage along a capillary of radius $r$, assumes that the average liquid viscosity remains constant along the radial direction of the flow and is not affected by the wall. Thomas \& McGaughey \cite{thomas2008reassessing} and Myers \cite{myers2011slip} proposed the slippage model with a variable viscosity at a distance away from the wall, the approach of which, to some extent, accounts for the liquid-solid interactions \cite{de2002fluid,hummer2001water,thomas2008reassessing}. A schematic of such models in a confined channel or pore is illustrated in Figure \ref{fig:Conceptual}.  Mattia \& Calabr\`o \cite{mattia2012explaining} further incorporated the surface diffusion and liquid adhesion near the wall surface, which characterizes the flow enhancement as a consequence of the migration of liquid molecules on the surface in addition to the viscous effect. The models reviewed here provide a parametrized approach for studying liquid transport in complex pore networks and, in particular, lay a theoretical basis for understanding shale oil transport.

Shale rocks are ultra-confined permeable media with a typical intrinsic permeability of less than 0.1 mD \cite{fan2016transient,fan2017analytical,rezaee2015fundamentals,song2019nonlinear}. The shale constituents are primarily divided into organic matter and inorganic minerals each having different wettability. The presence of organic and inorganic pores induces the characteristics of mixed wettability of shales. Inorganic clay minerals, e.g., kaolinite, illite, and smectite, are usually hydrophilic, while the organic matter, e.g., kerogen and bitumen, varies from highly hydrophobic to mixed-wet based on rock thermal maturity \cite{rezaee2015fundamentals}. Experimental study of oil and brine transport in mixed-wet limestones shows that the wetting phase can slip in mixed-wet rock, and that the slip length increases with a decreasing pore size \cite{christensen2017enhanced}. Recent studies attempted to model the apparent liquid permeability of shale rocks by applying the aforementioned flow enhancement models \cite{cui2019oil,cui2017liquid,fan2019enhanced,feng2019apparent,wang2019apparent,wang2019fractal,yang2019pore,zhang2017apparent}. Cui et al \cite{cui2017liquid} studied liquid slippage and adsorption in hydrophobic organic pores of shales and highlighted the importance of the adsorption layer for oil flow in organic pores of size 500 nm. Zhang et al. \cite{zhang2017apparent} modeled liquid slippage in inorganic pores and liquid adsorption in organic pores and showed that the wettability difference of these two pores leads to the fact that the apparent permeability of inorganic pores can be four orders of magnitude more than that of organic pores.

Differences in pore structures of inorganic versus organic matters with regards to pore size, pore size distribution, and surface roughness, also impact transport behaviors \cite{fan2017analytical}. The average size of organic pores is usually at least one order of magnitude less than that of inorganic pores. Organic pores are more uniformly distributed by size than inorganic ones, e.g., 18--438 nm for the former \cite{lu2015organic} versus 3 nm--100 $\mu$m for the latter \cite{chalmers2012characterization}. The surface roughness of pores is found scaled with pore sizes, e.g., the relative roughness, defined as the ratio of the roughness height divided by the local pore diameter, is often observed smaller in organic pores than inorganic ones \cite{javadpour2015slip}. The impact of surface roughness on transport is complex, e.g., slippage may be reduced due to stronger hydrogen bonding on rougher surfaces \cite{joseph2008carbon,myers2011slip} or enhanced due to the nano-scale `lotus effect' \cite{cao2006liquid}.

To date, studies on liquid transport behavior especially oil in mixed-wet porous media are limited. Understanding is still insufficient on the overall impact of pore structure and liquid-solid interactions of the rock permeability at the representative-elementary-volume (REV) scale, i.e., the smallest volume of which the measured permeability and porosity are statistically representative of, e.g., the whole rock core sample. This study develops a new apparent liquid permeability (ALP) model and provides an avenue for estimating the ALP of a chemical and spatially heterogeneous, confined permeable media via integrating atomistic and core-scale data. The data include core-flooding measurements, scanning electron microscope (SEM) images, mercury injection capillary pressure (MICP) tests, lattice Boltzmann (LB) simulations, MD simulations, and atomic force microscopy (AFM) results. The proposed ALP model presents the following contributions:
\begin{enumerate}
\item The ALP model quantifies liquid slippage contribution to the total flow rates on wetting and non-wetting surfaces.
\item The ALP model accounts for REV-scale heterogeneity in pore size and pore throat tortuosity, and pore-scale roughness on liquid slippage.
\item The ALP model compiles MD data readily via detailed workflows proposed in this work.
\item The ALP model clarifies the analogies and differences between the shale permeability model and classic permeability models. In particular, a critical comparative analysis between the proposed ALP model and the Carman, the Carman-Kozeny, and the Klinkenberg equations is presented to highlight the virtues and features of the ALP model. 
\end{enumerate}

\section{Method}
This section presents the derived liquid slippage and the ALP for heterogeneous, tortuous, rough, and mixed-wet porous media at the REV scale. The ALP combines the effect of pore structure, near-wall flow regions as well as fluid-rock interactions.
\subsection{Flow enhancement model}
In shale rocks, properties of near-wall regions are different in inorganic and organic pores due to wettability. In addition to ``free oil'', organic pores, typically hydrophobic ones, are rich in adsorbed oil \cite{ambrose2010new}. The flow in cylindrical organic pores accordingly can be divided into two viscous regions: a cylindrical bulk flow region of viscosity $\mu_b$ and an annular near-wall region of thickness $\delta_w$ and viscosity $\mu_w$. The concept of this bi-viscosity model also applies to hydrophilic inorganic pores because the strong hydrophilicity promotes the oil slippage within such pores, rendering the viscosity near the wall lower than the bulk value \cite{wang2016molecular}.

\begin{figure}
\centering
\includegraphics[width=\textwidth]{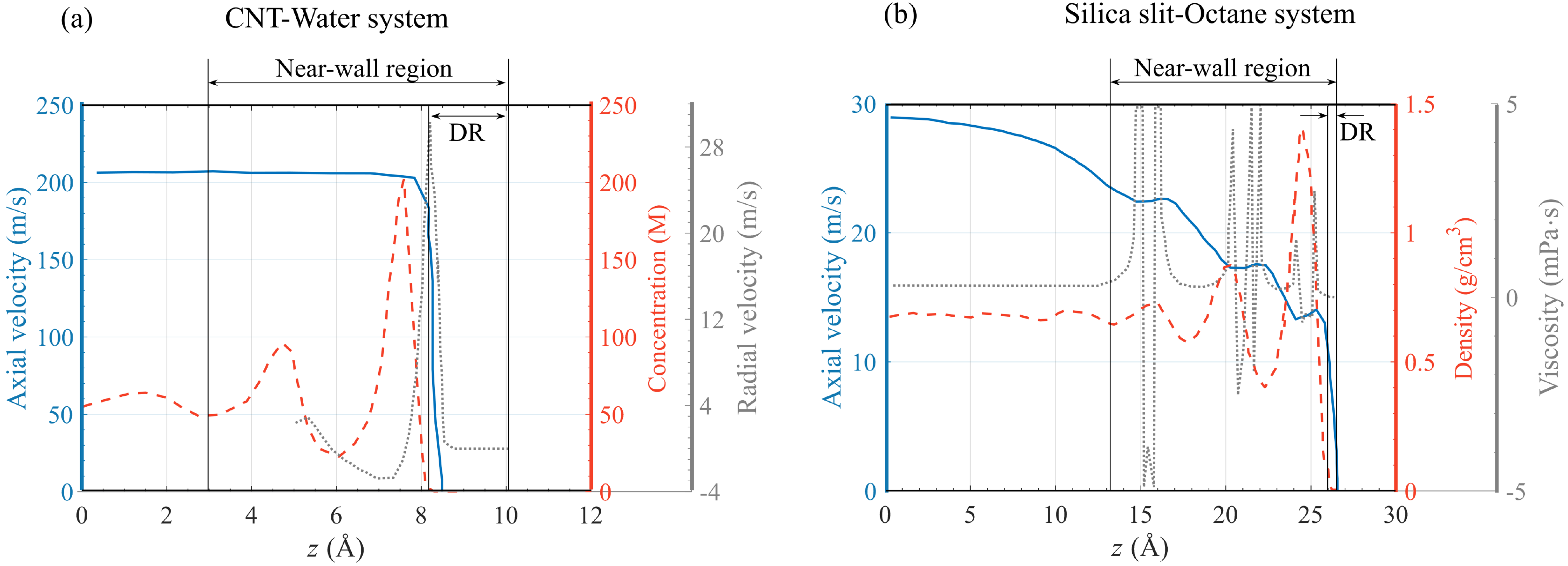}
\caption{MD data for (a) water transport in a 2.17-nm CNT \cite{joseph2008carbon} and (b) octane transport in a 5.24-nm silica slit \cite{RN23}. The information in (b) will be discussed in Section \ref{sec:aaa}. ``DR'' denotes the depletion region. } 
\label{fig:NearwallRegion}
\end{figure}

Of note, the ``near-wall region'' here refers to as the region where the local fluid density deviates from the bulk density from the pore surface. This region should not be confused with the ``depletion region (DR)'', the formation of which, e.g., for water on hydrophobic surfaces, is due to the repulsive electrostatic interactions between water molecules and nonpolar surfaces, and this region typically refers to the region where the local liquid density is less than 2\%--5\% of the bulk density \cite{joseph2008carbon,RN14}. To clarify the definition of the two regions, we present an example of water flow through carbon nanotubes (CNTs) in Figure \ref{fig:NearwallRegion}(a). At the DR, water concentration decreases intensively, which corresponds to ``velocity peak'' and ``velocity jump'' in radial and axial velocity \cite{joseph2008carbon,RN22}. This region is $\sim$2 \r{A} thick, close to the value reported in the literature as one water molecule layer, i.e., $\sim$2.75 \r{A} \cite{RN17,RN15,RN19,RN18,RN16}. In contrast, the ``near-wall region'' is much wider, e.g., $\sim$7 \r{A} \cite{joseph2008carbon}. Similar rules to identify the near-wall region and the DR also apply to the silica-octane system (Figure \ref{fig:NearwallRegion}(b)).

Following the methodology described by \cite{mattia2012explaining}, we derive the intrinsic volumetric flow rate $Q$ (Equation \eqref{eq:q}) and the apparent volumetric flow rate $Q_{app}$ (Equation \eqref{eq:Qapp}), in which the Ruckenstein's slip (Equation \eqref{eq:lslipr}) is used to account for the contribution of surface diffusion and liquid adhesion to flow enhancement. The ratio of $Q_{app}$ and $Q$ yields the flow enhancement factor \cite{mattia2012explaining}:
\begin{equation}\label{eq:f}
	f = (\frac{\mu_b}{\mu_w}-1)(1-\lambda_b^2)+\lambda_s,
\end{equation}
where 
\begin{equation}\label{eq:lambdab}
	\lambda_b = \Big(1-\frac{\delta_w}{r_{p}}\Big)^2
\end{equation}
is the pore-structure factor and 
\begin{equation}\label{eq:lambdas}
\lambda_{s} = \frac{8\mu_bD_{s}L_{s}}{r_{p}^2W_{A}}+1
\end{equation}
is the slippage factor \cite{thomas2008reassessing,mattia2012explaining}. $r_p$ is the pore radius. $L_s$ is the straight pore length. $W_A$ is the work of adhesion which quantifies the energy of liquid adhesion per solid surface area. 

The presented $\lambda_s$ (Equation \eqref{eq:lambdas}) has readily accounted for the effect of pore size and other transport properties, such as viscosity and surface diffusion, yet the impact of pore throat tortuosity and surface roughness is not quantified. To include tortuosity, we substitute $L_s$ with a tortuous pore length $L_p$. The relation of $L_s$ and $L_p$ is evaluated by the (diffusive) tortuosity \cite{ghanbarian2013tortuosity}:
\begin{equation}\label{eq:torttt}
    \tau=\Big(\frac{L_p}{L_s}\Big)^2.
\end{equation}
By introducing a tortuosity fractal dimension $D_T$, we describe pore throat as fractals and the $L_p$ is estimated by Equation \eqref{eq:Lp}. To include the roughness effect, we recall a fractal relative roughness $\varepsilon$ in Equation \eqref{eq:e}. By recalling Equations \eqref{eq:Lp} and \eqref{eq:e}, the formulation of apparent liquid slippage is derived:
\begin{equation}\label{eq:lambdasapp}
\lambda_{s,app} =  \Big[\frac{2^{-D_{T}+4}\mu_bD_sL_{s}^{D_{T}}}{(\overline{r}_{p})^{D_{T}+1}W_A} + 1\Big](1-\varepsilon)^4,
\end{equation}
for a pore of the average radius ($\overline{r}_{p}$) weighted averaged over the REV, where $
\overline{r}_{p}=\overline{d}_{p}/2= -\int_{d_{p,min}}^{d_{p,max}} d_{p}\mathbf{d}N_{p}(d_{p})/2N_{t}= (d_{p,min}-\gamma^{D_{p}}d_{p,max})D_{p}/(2D_{p}-2)$. $N_t$ is the total number of pores in an REV, estimated by $\gamma^{-D_{p}}$; $\gamma = d_{p,min}/d_{p,max}$ is the pore-size heterogeneity coefficient; $D_p$ is the pore size fractal dimension. 

Equation \eqref{eq:lambdasapp} is an important modification to Equation \eqref{eq:lambdas} because the former allows one to quantify liquid slippage through a realistic pore structure, which is typically non-straight and rough. When $D_T=1$ and $\varepsilon=0$, Equation \eqref{eq:lambdasapp} reduces to Equation \eqref{eq:lambdas} for a straight and smooth pore. 

\subsection{Apparent liquid permeability (ALP)}
The intrinsic permeability is derived by recalling Equation \eqref{eq:k1} and the fractal relations of pore size (Equation \eqref{eq:N_p}), pore throat tortuosity (Equation \eqref{eq:Lp}), and pore-surface roughness distributions (Equation \eqref{eq:N_c}) in an REV: 
	\begin{equation}\label{eq:k2}
		k= \frac{d_{p,max}^2}{32}\frac{\phi}{\tau(d_{p,max})}\xi(D_T, D_p, \varepsilon, \gamma),
	\end{equation}
where 
\begin{equation}
    \xi(D_T, D_p, \varepsilon, \gamma)=\frac{(-D_T-D_p+3)(1-\varepsilon)^4}{(D_{T}-D_{p}+3) (1-\gamma^{-D_T-D_p+3})}.
\end{equation}
is the fractal function that embraces surface roughness and pore size distribution information. $d_{p,max}$ is the maximum pore diameter in the REV. $\phi = \gamma^{3-D_p}$ is the fractal porosity \cite{Wei2015An,yu2008analysis}. $\tau(d_{p,max})= (d_{p,max}/L_s)^{-2D_T+2}$ is the tortuosity of the maximum pore diameter, derived by combining Equations \eqref{eq:torttt} and \eqref{eq:Lp}. Relevant pore-scale fractal models are presented in Appendix \ref{sec:fractal}.

Combining Equations \eqref{eq:f} and \eqref{eq:k2} yields the ALP: 
\begin{equation}\label{eq:kapp2}
	\begin{aligned}
		k_{app}&= k\times f \\
		& = \Big\{(\frac{\mu_b}{\mu_w}-1)\Big[1-(1-\frac{\delta_w}{\overline{r}_{p}})^4\Big]+\lambda_s\Big\}\\
		&\times\frac{d_{p,max}^{2D_{T}}}{32 L_{s}^{2D_{T}-2}}
		\frac{(-D_{T}-D_{p}+3)}{(D_{T}-D_{p}+3) }\frac{\gamma^{-D_{p}+3}}{(1-\gamma^{-D_{T}-D_{p}+3})}(1-\varepsilon)^4\\
		&= \Big\{(\frac{\mu_b}{\mu_w}-1)\Big[1-(1-\frac{\delta_w}{\overline{r}_{p}})^4\Big]+\Big[\frac{2^{-D_{T}+4}\mu_bD_sL_{s}^{D_{T}}}{(\overline{r}_{p})^{D_{T}+1}W_A} + 1\Big]\Big\}\\
&\times\frac{d_{p,max}^{2D_{T}}}{32 L_{s}^{2D_{T}-2}}
\frac{(-D_{T}-D_{p}+3)}{(D_{T}-D_{p}+3) }\frac{\gamma^{-D_{p}+3}}{(1-\gamma^{-D_{T}-D_{p}+3})}(1-\varepsilon)^4.\\
	\end{aligned}
\end{equation}
where $\lambda_s$ is substituted with $\lambda_{s,app}/(1-\varepsilon)^4$. The derived ALP model is then applied to estimate apparent permeability in inorganic matters ($k_{i,app}$) and organic matters ($k_{o,app}$).

\subsection{ALP estimation workflow}\label{sec:workflow}
\begin{figure}[t]
\centering
\includegraphics[height=0.8\textwidth]{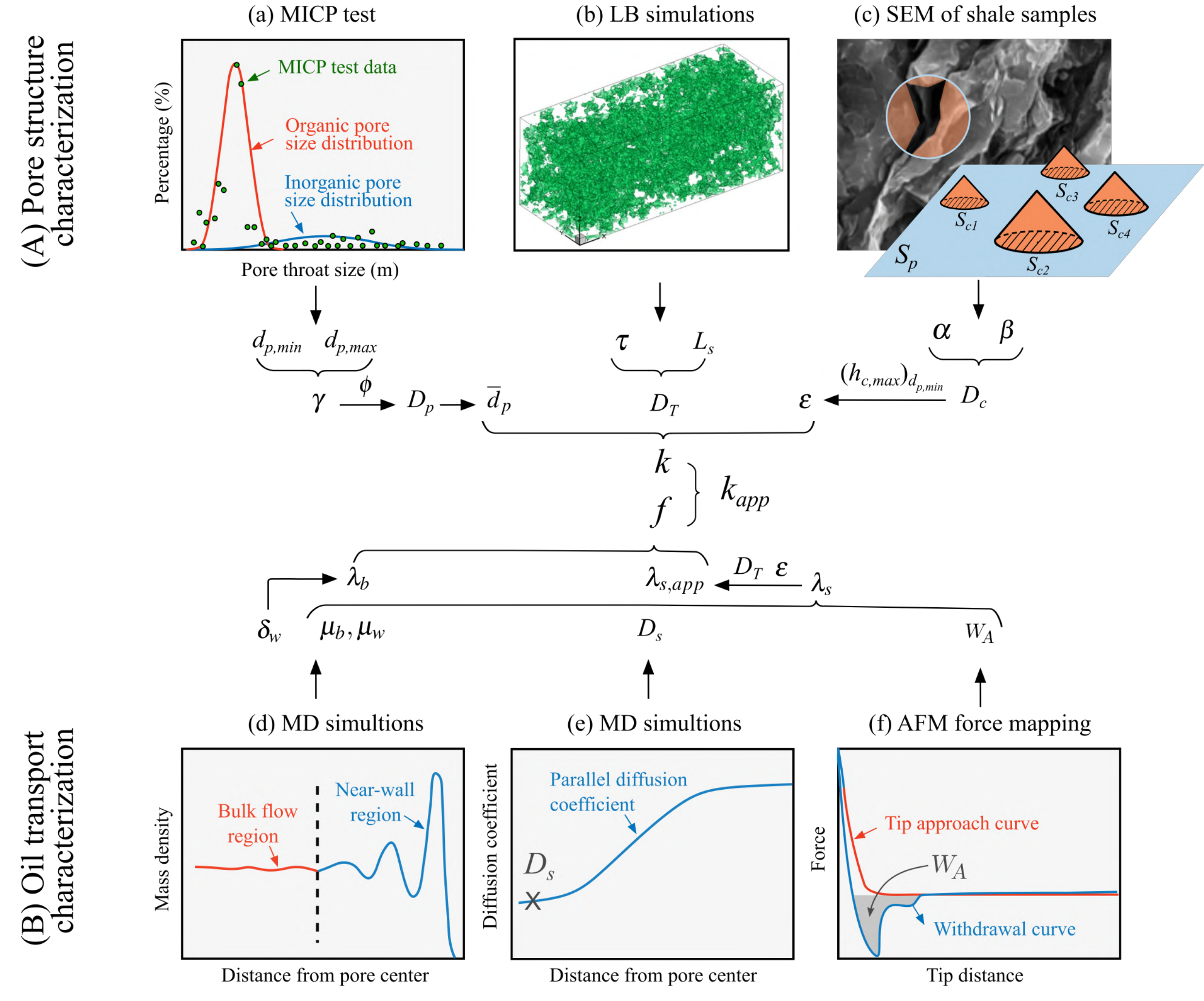}
\caption{Flowchart of the ALP model for nanoporous shales \cite{fan2019confinement,fan2018fluids}. (a-c) Extraction of pore structure information, i.e., pore size, tortuosity, and surface roughness via of MICP experiments, LB simulations \cite{chen2015nanoscale}, and SEM images \cite{yang2015new}, respectively. (d-f) Quantification of oil transport properties, i.e., near-wall region thickness and viscosity, surface diffusion, and work of adhesion via MD simulations and AFM force mapping, respectively. Intrinsic permeability ($k$) and flow enhancement factor ($f$) are coupled to estimate ALP ($k_{app}$).}. 
\label{fig:Workflow}
\end{figure}
\begin{table}[t]
	\centering
	\caption{Input data for the ALP model, aggregated from literature, and compiled.}
	\label{tab:inputs}
	\small
	\begin{threeparttable}  
		\begin{tabular}{l|l|l}
			\hline
			Properties & Values (SI unit) & References\\ \hline
			Inorganic matrix& &\\
			$d_{pi,max}$& 1.63$\times10^{-5}$ m & \text{\cite{josh2012laboratory}}\\
			$d_{pi,min}$& 2.30$\times10^{-8}$ m & \cite{josh2012laboratory} \\			
			$\gamma_i$ & 1.42$\times10^{-3}$ & $\gamma_i = d_{pi,min}/d_{pi,max}$\\
			$\phi_i$ & 0.06 & \cite{fan2017analytical}\\
			$D_{pi}$ & 2.57 & $D_{pi} = 3-ln(\phi_i)/ln(\gamma_i)$\\
			$\tau_i$ & 66 & $\tau_i = \phi_i^{-1.49}$ in \cite{chen2015nanoscale}\\
			$D_{Ti}$ & 1.38 & $D_{Ti} = 1-ln(\tau_i)/2ln(\overline{d}_{pi}/L_{si})$\\
			$\beta_i$ & 0.02 & \cite{yang2014fractal}  \\
			$\alpha_i$ & 0.002 & $\alpha_i = \beta_i^{3-D_{ci}}$\\
			$(h_{ci,max})_{d_{pi,min}}/d_{pi,min}$ & 0.50 & SEM images in \cite{yang2015new} \\
			$D_{ci}$ & 1.40 & \cite{yang2015permeability} \\
			$L_{si} = d_{mi}$ & 1$\times10^{-5}$ m & \cite{chen2015permeability} \\
			$\mu_b$ & 9.6$\times10^{-4}$ Pa$\cdot$s & \cite{whitby2008enhanced}\\
			$\mu_b/\mu_w$ & 1.67 & \cite{mattia2012explaining}\\ 
			$\delta_w$ & 7$\times10^{-10}$ m & \cite{joseph2008carbon,mattia2012explaining}\\
			$D_{si}$ & 3$\times10^{-9}$ m$^2$/s & \cite{mattia2012explaining}\\ 
			$W_{Ai}$ & 0.025 J/m$^2$& \cite{hassenkam2009probing}\\ \hline
			Organic matrix & & \\
			$d_{po,max}$& 8.88$\times10^{-8}$ m & \cite{josh2012laboratory}\\
			$d_{po,min}$& 3.84$\times10^{-9}$ m & \cite{josh2012laboratory}\\	
			$\gamma_o$ & 4.32$\times10^{-2}$ & $\gamma_o = d_{po,min}/d_{po,max}$\\
			$\phi_o$ & 0.03 & \cite{fan2017analytical}\\
			$D_{po}$ & 1.88 & $D_{po} = 3-ln(\phi_o)/ln(\gamma_o)$\\
			$\tau_o$ & 4518 & $\tau_o = \phi_o^{-2.40}$ in \cite{chen2015pore} \\
			$D_{To}$ & 1.86 & $D_{To} = 1-ln(\tau_o)/2ln(\overline{d}_{po}/L_{so})$\\
			$\beta_o$ & 0.02 &\cite{yang2014fractal}\\
			$\alpha_o$ & 0.001 & \cite{javadpour2015slip}\\
			$(h_{co,max})_{d_{po,min}}/d_{po,min}$ &0.05 & SEM images in \cite{javadpour2015slip,yang2015permeability}\\
			$D_{co}$ & 1.23 & $D_{co} = 3-ln(\alpha_o)/ln(\beta_o)$\\ 
			$L_{so} = d_{mo}$ & 1$\times10^{-6}$ m & \cite{chen2015permeability} \\ 
			$\mu_b/\mu_w$ & 0.91 & \cite{zhang2017apparent}\\ 
			$\delta_w$ & 1$\times10^{-9}$ m & \cite{wang2015oil}\\ 
			$D_{so}$ & 1$\times10^{-9}$ m$^2$/s & \cite{ershov2001displacement}\\ 
			$W_{Ao}$ & 0.144 J/m$^2$& \cite{mattia2012explaining}\\ \hline
		\end{tabular} 
	\end{threeparttable}  
	\end{table}
To demonstrate the use of the ALP model, we here provide a workflow to estimate ALP parameters using lab experimental and MD results. Table \ref{tab:inputs} summarizes the data for the key fractal and transport parameters reported in the literature. Figure \ref{fig:Workflow} illustrates this workflow, summarized in three major steps:
\begin{enumerate}[label=Step \arabic*.]
\setlength{\itemindent}{0.2in}
    \item Quantify pore structure to calculate $k$ in Equation \eqref{eq:k2}.
    \item Quantify liquid transport, where we model the bulk flow region, the near-wall region, and strength of liquid-solid interactions to calculate $f$ in Equations \eqref{eq:f} through \eqref{eq:lambdasapp}.
    \item Couple Steps 1 and 2 to derive $k_{app}$ in Equation \eqref{eq:kapp2}.
\end{enumerate}
	
\paragraph{Pore size distribution (PSD).} \label{sec:pore}
Mercury injection capillary pressure (MICP) test is classically used to estimate PSDs \cite{josh2012laboratory}. Figure \ref{fig:Workflow}(a) shows the MICP results of a bimodal PSD for a shale sample. Following the methodology by \cite{naraghi2015stochastic}, the bimodal PSD is divided into two distributions: a widely spread inorganic distribution and a narrowly spread organic distribution. The distributions are parameterized by the pore size fractal relation in Equation \eqref{eq:N_p}. 

\paragraph{Pore throat tortuosity.}
Tortuosity data are usually acquired by flow simulations, e.g., lattice Boltzmann (LB) simulations (Figure \ref{fig:Workflow}(b)). The (diffusive) tortuosity (Equation \eqref{eq:torttt}) is found to obey an empirical, power-law scaling law with porosity, i.e., the Bruggeman's equation: $\tau = \phi^{-n}$ \cite{chen2015nanoscale,shen2007critical}. This scaling law allows us to acquire the estimated $\tau$ from the flow simulation results. Once $\tau$ is obtained, we are able to parametrize pore throat fractals, e.g., $D_T$. The exponent $n$ is empirical and varies with pore structures of different samples. For high-porosity media, $n$ was estimated $\sim0.5$ \cite{shen2007critical}. For low-porosity shale samples, the LB simulations yielded $n$ = 1.33--1.65 for shale bulk \cite{chen2015nanoscale} and 1.8--3 for organic matters in shale \cite{chen2015pore}. We accordingly adopt the average $n$ to calculate tortuosity for inorganic pores ($\tau_i$) and organic pores ($\tau_o$): $\tau_i = \phi_i^{-1.49} = 66$ and $\tau_o = \phi_o^{-2.40} = 4518$, where porosity $\phi_i = 0.06$ and $\phi_o = 0.03$, as summarized in Table \ref{tab:inputs}. The estimated tortuosity values are in the typical range of shale samples reported from the literature, i.e., 100--1000 \cite{RN1,ghanbarian2017upscaling,woodruff2015measurements}. 

\paragraph{Pore-surface roughness.}
Equation \eqref{eq:e} is used to estimate the relative roughness on pore surfaces. Figure \ref{fig:Workflow}(c) is an example SEM image of inorganic matters \cite{yang2015new} in a shale sample; it also illustrates the schematic of modeling surface roughness as many conical nanostructures inside the pore as well as shows the distribution of those nanostructures if the pore surface ``spreads out'' as a plane. Key parameters such as the areal ratio ($\alpha$) and conical height ($(h_c)_{d_p}$) in Equation \eqref{eq:e} are estimated via length and areal calculations of the structures observed in the SEM images. The fractal dimension of the conical base size distribution ($D_c$) is estimated via interpreting conical base size and number of cones and using Equation \eqref{eq:N_c}. 
	
\paragraph{Near-wall region.} \label{sec:deee}
Figure \ref{fig:Workflow}(d) shows the MD results of octane density in an inorganic pore \cite{RN23}. From the density fluctuation, we identify the near-wall region and the bulk flow region. Based on the MD results, the thickness fraction of the near-wall region ($\delta_w/r_{pi}$) and the factor $\lambda_{bi}$ are estimated by Equation \eqref{eq:lambdab}. A similar procedure is conducted for estimating parameters of octane transport through an organic pore.

\paragraph{Surface diffusion, work of adhesion, slippage, \& flow enhancement.}
The surface diffusion coefficient ($D_s$) is derived from MD simulations by evaluating the self-diffusion coefficient parallel with the wall in which the coefficient in the first molecular layer is adopted as the value of $D_s$, as shown in Figure \ref{fig:Workflow}(e). The work of adhesion is obtained via atomic force microscopy (AFM) mapping results. Figure \ref{fig:Workflow}(f) presents the AFM map of force versus distance for tip approach and withdrawal  \cite{hassenkam2009probing,argyris2011structure,kobayashi2016molecular}, where the encompassed gray area estimates the work of adhesion $W_{A}$. The apparent slippage factor ($\lambda_{s,app}$) is estimated by $\mu_b$, $\mu_w$, $D_s$, $W_A$, $D_T$, and $\varepsilon$. The flow enhancement factor ($f$) is calculated based on $\lambda_{s,app}$ and $\lambda_b$.

\paragraph{Literature data for confined oil transport.}
We review some literature data of key transport properties of hydrocarbon liquids on hydrophilic and hydrophobic surfaces in Table \ref{tab:MDdatasummary}. Through examining Table \ref{tab:MDdatasummary}, we find: (1) A wide range of slip length of octane has been reported, i.e., 0 to $>$130 nm in different MD studies, implying a strong dependence of liquid slippage on the substrate type, driving force, and substrate surface roughness. (2) The total near-wall region thickness ($2\delta_w$) of octane is found dependent on the pore confinement: In narrow hydrophilic slits, e.g., $H=$ 2 nm, the fluctuation of near-wall viscosity may not stabilize at the slit center, which diminishes the bulk region and cause $2\delta_w/H\rightarrow 1$; In narrow hydrophobic slits, e.g., $H<$ 3.9 nm, the bulk-density may not present, which is due to the superimposition of the interaction potentials as well as the adsorption layers from substrate surfaces. Compared to the near-wall thickness in hydrophilic slits, total adsorption layer thickness fraction in hydrophobic slits is more consistent for 2--5-nm slits, i.e., around 40\%--50\% of the entire flow region. (3) $W_A$ and $D_s$ data for octane are generally limited in the current body of the literature. 

\section{Results}\label{sec:factorsoil}
\subsection{Validation against MD data}
Recent ALP models on liquid slippage in shale matrices have shown their ability to predict the enhancement of the confined water transport in straight nanotubes via MD data, yet their capability of predicting liquid (including oil and water) transport in tortuous and rough nanopores is unknown \cite{cui2017liquid,fan2019enhanced,zhang2017apparent,RN35}. Here, we validate our model against a series of MD data in the literature for 
\begin{enumerate}
    \item confined octane transport in straight slit pores, estimated by Equation \eqref{eq:lslipH};
    \item confined liquid transport in tortuous cylindrical pores, estimated by Equation \eqref{eq:lll};
    \item confined liquid transport in rough cylindrical pores, estimated by Equation \eqref{eq:lambdasapp}.
\end{enumerate}
\begin{figure}[t]
	\centering
	\includegraphics[width=\linewidth]{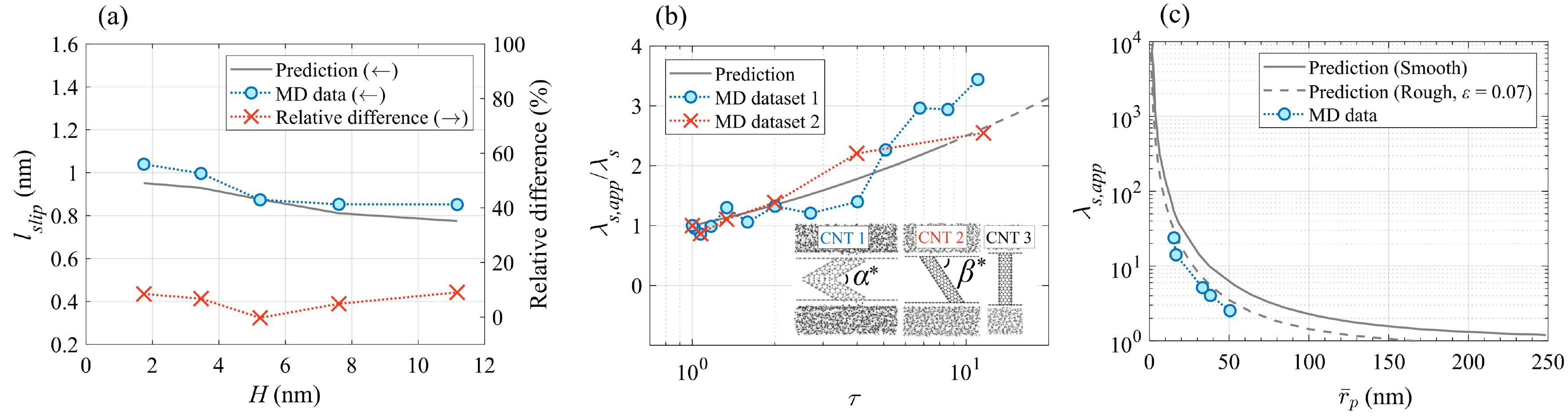}
	\caption{(a) Comparison of the MD data \cite{RN23} for octane transport in a silica slit versus predictions from Equation \eqref{eq:lslipH}. Relative differences are shown for prediction deviations from the MD data. (b) Comparison of the MD data for tortuous nanotubes versus $\lambda_{s,app}$ predictions from Equation \eqref{eq:lll}. MD dataset 1 \cite{RN42} and dataset 2 \cite{RN43} correspond to CNT Type 1 with a bending angle $\alpha^*$ and Type 2 with a tilting angle $\beta^*$, respectively. Different tube tortuosity is achieved via alternating $\alpha^*$ and $\beta^*$; Tortuous length and tube size are fixed as $L_p =$ 3.8 nm and $d_p=$ 0.777 nm of CNT Type 1; and $L_p=$ 3.824 nm and $d_p=$ 0.782 nm of CNT Type 2. Straight CNT configuration is shown as CNT Type 3. Temperature $T =$ 300 K; bulk viscosity $\mu_b\approx$ 0.85 mPa$\cdot$s \cite{kestin1984thermophysical}; work of adhesion $W_A =$ 97 mJ/m$^2$; surface diffusion coefficient $D_s$= 4$\times10^{-9}$ m$^2$/s \cite{mattia2012explaining}. (c) Comparison of the MD data for a rough nanotube \cite{secchi2016massive}, a slip model for smooth CNTs \cite{zhang2017apparent}, versus $\lambda_{s,app}$ predictions from Equation \eqref{eq:lambdasapp}.}
	\label{fig:MD}
\end{figure}
\subsubsection{Confined octane transport in straight slit pores} \label{sec:aaa}
In prior studies, the Ruckenstein's slip (Equation \eqref{eq:lslipr}) was applied to confined water flow through CNTs \cite{mattia2012explaining}. Recently proposed ALP models \cite{cui2017liquid,fan2019enhanced,zhang2017apparent,RN35} assumed that Equation \eqref{eq:lslipr} is capable of describing the slip length of oil flows; however, direct MD validations are lacking. Indeed, Equation \eqref{eq:lslipr} is the basis of Equation \eqref{eq:lambdasapp}, of which the latter is an important ingredient of the ALP model proposed in this work. We revisit Ruckenstein's slip model to investigate whether its theory can describe the oil slippage. For slit pore configurations, the Ruckenstein's slip model is corrected as \cite{RN38}: 
\begin{equation} \label{eq:lslipH}
    l_{slip} = \frac{2\mu_wL_sD_s}{HW_A}.
\end{equation}

\begin{table}[!]
\centering
\caption{MD data \cite{RN23} for octane transport through a 5.24 nm silica slit.}
\label{tab:MDdatavalidation}
\begin{tabular}{lll}
\hline& Property  & Value\\ \hline
\multirow{5}{*}{Input} & Slit length  $L_s$ (nm)   & 2.9  \\
& Slit aperture $H$ (nm)  & 5.24  \\
& Surface diffusion coefficient $D_s$ (m$^2$/s)   & 2.88$\times10^{-9}$\\
& Bulk viscosity $\mu_b$ (mPa$\cdot$s)& 0.359 \\
& Effective viscosity $\mu_{eff}$ (mPa$\cdot$s)  & 0.295   \\ \hline
\multirow{4}{*}{Output} & Slip length $l_{slip}$ (nm) from Step 1 & 0.874   \\
& Near-wall thickness $\delta_w$ (nm) from Step 2  & 1.26  \\
& Near-wall viscosity $\mu_w$ (mPa$\cdot$s) from Step 3 & 0.226 \\
& Work of adhesion $W_A$ (J/m$^2$) from Step 4 & 8.24$\times10^{-4}$ \\ \hline
\end{tabular}
\end{table}

To validate Equation \eqref{eq:lslipH} for shale oil transport, we compile the MD data in Table \ref{tab:MDdatavalidation} for octane flow through a straight, silica slit \cite{RN23}. The velocity profile is presented in Figure \ref{fig:NearwallRegion}(b). In Equation \eqref{eq:lslipH}, we assume that:
\begin{enumerate}
    \item $L_s$ is the length of the slit in the axial direction.
    \item Slit confinement has little impact on the liquid adsorption, i.e., $W_A$ is independent of $H$.
    \item The values of $D_s$ and $\mu_w$ vary with $H$ (according to MD simulation results \cite{RN39,RN40,RN26}).
\end{enumerate}

The following algorithm is implemented to estimate $l_{slip}$ for different slit apertures:
\begin{enumerate}[label=Step \arabic*.]
\setlength{\itemindent}{0.2in}
    \item Estimate $l_{slip}$ from the MD velocity profile for the 5.24-nm slit. The value of $l_{slip}$ is estimated by extrapolating the MD velocity beyond the liquid-solid interface until the liquid velocity vanishes, where $l_{slip}=-v_{slip}/(\mathbf{d}v/\mathbf{d}z)_{wall}$, $v_{slip}$ is the slip velocity at the wall, $z$ is the direction perpendicular to the wall \cite{RN41}. 
    \item Estimate $\delta_w$ based on the MD density profile for the 5.24-nm slit (Figure \ref{fig:NearwallRegion}(b)). 
    \item Estimate $\mu_w$ for the 5.24-nm slit from the MD viscosity profile via either of the following methods. One can estimate $\mu_w$ via averaging the liquid viscosity in the identified the near-wall region from Step 2. An alternative method is to calculate $\mu_w$ from the effective viscosity data ($\mu_{eff}$) if the latter is available. The effective viscosity is the weighted average based on the fraction of the cross-sectional areas of the bulk flow and the near-wall region, i.e., Equation \eqref{eq:visco}:  $\mu_{eff}=\mu_w A_w/A_t+\mu_b (1-A_w/A_t)$ where $A_w=2\delta_w L$ and $A_t=HL$ are the cross-sectional area of the near-wall region of thickness $\delta_w$ and the entire flowing region in an $H$-aperture slit, respectively. In this way, 
    \begin{equation}
        \mu_{w}=\frac{H}{2 \delta_{w}}\left[\mu_{e f f}-\mu_{b}\left(1-\frac{2 \delta_{w}}{H}\right)\right].
    \end{equation}
    \item Estimate $W_A$ by Equation \eqref{eq:lslipH}.
    \item Repeat Step 1 for slit aperture $H$= 1.74 nm, 3.46 nm, 7.61 nm, and 11.17 nm.
    \item Estimate $D_s$ for different slit apertures. In the literature, the self-diffusion coefficient ($D_{self}$) of water, n-octane, octanol, dimethyl sulfoxide as well as supercritical methane were found to increase with confinement when $H\lesssim$ 10 nm, the relation of which can be described in a linear function \cite{RN39,RN40,RN26}. For all cases studied in Step 5, slit apertures are $<$ 12 nm; we assume that the linearity holds for the surface diffusion coefficient $D_s$. Given $W_A$, $\delta_w$, $l_{slip}$ in Table \ref{tab:MDdatavalidation} we estimate $D_s$ ($H =$ 7.61 nm) = 4.24$\times10^{-9}$ m$^2$/s. Now with $D_s$ ($H =$ 5.24 nm), the linear function of $D_s = 0.57H-0.1$ is obtained, where $H$ is in nm and $D_s$ is in 1$\times10^{-9}$ m$^2$/s. A series of $D_s$ for $H\leq$12 nm is estimated accordingly.
    \item Repeat Steps 2 and 3 to estimate $\delta_w$ and $\mu_w$ for different slit apertures based on their density profiles. Density data can be found in \cite{RN23}. 
    \item Calculate $l_{slip}$ for different slit apertures by Equation \eqref{eq:lslipH}.  
\end{enumerate}
In Figure \ref{fig:MD}(a), the estimated $l_{slip}$ from Step 8 is plotted against MD predictions from Step 1. Although slightly overestimating the slip length (possibly due to the simplified approximation of $D_s$ values), Equation \eqref{eq:lslipH} generally captures the octane slippage in silica slits of  $H \leq$ 12 nm, given the small difference, i.e., $\leq 9\%$, observed between MD data and predictions. Direct MD data of $D_s$ for different apertures can improve the $l_{slip}$ predictions when using Equation \eqref{eq:lslipH}.  

\subsubsection{Confined liquid transport in tortuous cylindrical pores}
We propose Equation \eqref{eq:lambdasapp} to estimate apparent liquid slippage on tortuous, rough cylindrical nanopores. Assuming that the impact of surface roughness on liquid slippage is negligible (with $\varepsilon \rightarrow 0$) when compared to the impact of tortuosity, Equation \eqref{eq:lambdasapp} reduces to
\begin{equation}\label{eq:lll}
    \lambda_{s,app} = \frac{2^{-D_{T}+4}\mu_bD_sL_{s}^{D_{T}}}{(\overline{r}_{p})^{D_{T}+1}W_A} + 1.
\end{equation}

To testify the ability of Equation \eqref{eq:lll} to predict liquid slippage in tortuous pores, we adopt MD data for confined water transport in bent and tilted CNTs \cite{RN42,RN43}. For such geometry, Equation \eqref{eq:torttt} is further extended in terms of trigonometric ratios:
\begin{equation}\label{eq:ttt}
    \tau=\Big(\frac{L_p}{L_s}\Big)^2=\Big[\sin\Big(\frac{\alpha^*}{2}\Big)\Big]^{-2}=\Big[\sin(\beta^*)\Big]^{-2},
\end{equation}
where $\alpha^*$ and $\beta^*$ are bending and tilting angles, respectively, as illustrated in Figure \ref{fig:MD}(b).

To ensure that $D_T$ is physically meaningful, i.e., $1\leq D_T \leq 3$, tortuosity should be $1\leq \left[\tau=(d_p/L_s)^{-2D_T+2} \right]\leq (d_p/L_s)^{-4}$. The validity of Equation \eqref{eq:lll}, therefore, depends on the pore size distribution of the studied samples. For example, when $d_p/L_s>1$, Equation \eqref{eq:lll} is not applicable; when $d_p/L_s =0.4$, Equation \eqref{eq:lll} is applicable if tortuosity is $1\leq \tau \leq 36$. Given the data for $d_p$ and $L_p$ \cite{RN42,RN43} as well as the requirement of $1\leq D_T\leq 3$, Equation \eqref{eq:lll} holds when the bending angle $\alpha^* \geq \frac{360}{\pi} \arcsin [(\frac{d_p}{L_p})^{\frac{2}{3}}] =40.617^{\circ}$ and $\beta^* \geq \frac{180}{\pi} \arcsin[(\frac{d_p}{L_p})^{\frac{2}{3}}]=20.309^{\circ}$. These angles correspond to $\tau\leq$ 8.3.

The following algorithm is performed to validate Equation \eqref{eq:lll}:
\begin{enumerate}[label=Step \arabic*.]
\setlength{\itemindent}{0.2in}
    \item Estimate $\tau$ via $\alpha^*$ or $\beta^*$ data and Equation \eqref{eq:ttt}.
    \item Estimate $L_s$ for different angles: $L_s=L_p\sin(\alpha^*/2)=L_p\sin(\beta^*)$.
    \item Estimate $D_T$ via $\tau$: $D_T=1-\frac{ln(\tau)}{2ln(d_p/L_s)}$.
    \item Estimate $\lambda_s$ for straight CNT (Type 3) by Equation \eqref{eq:lambdas}.
    \item Estimate $\lambda_{s,app}$ for tortuous CNTs (Types 1 and 2) by Equation \eqref{eq:lll} with inputs of $D_T$ from Step 3. 
\end{enumerate}

Figure \ref{fig:MD}(b) compares the prediction results from Steps 4 and 5 against the MD data. Detailed descriptions of the dataset are captioned in Figure \ref{fig:MD}(b). The result demonstrates that Equation \eqref{eq:lll} is a good predictor of the liquid slippage in tortuous pores. 

\subsubsection{Confined liquid transport in rough cylindrical pores}
Another feature of Equation \eqref{eq:lambdasapp} is the consideration of surface roughness. In Equation \eqref{eq:lambdasapp}, roughness is modeled as resistance to liquid slippage. To validate this resistance effect, we adopt the MD data for confined water in straight, rough CNTs \cite{secchi2016massive}. 
Figure \ref{fig:MD}(c) compares the MD data \cite{secchi2016massive}, the apparent slippage factor estimated by Equation \eqref{eq:lambdasapp} (with $D_T=1$ for the straight CNT), and a slippage model for smooth CNTs \cite{zhang2017apparent}. The results show that the apparent slippage factor predicted by Equation \eqref{eq:lambdasapp} agrees with the MD results better than the prior model for smooth CNTs \cite{zhang2017apparent}, which highlights the importance of the surface roughness on liquid slippage. Relative roughness is estimated to be $\varepsilon =$ 0.07 through data matching. 
\subsection{Governing factors of confined oil transport}
This section presents the analysis results for the underlying factors that control oil transport in shale rocks. 

\begin{figure}[t]
	\centering
	\includegraphics[width=\linewidth]{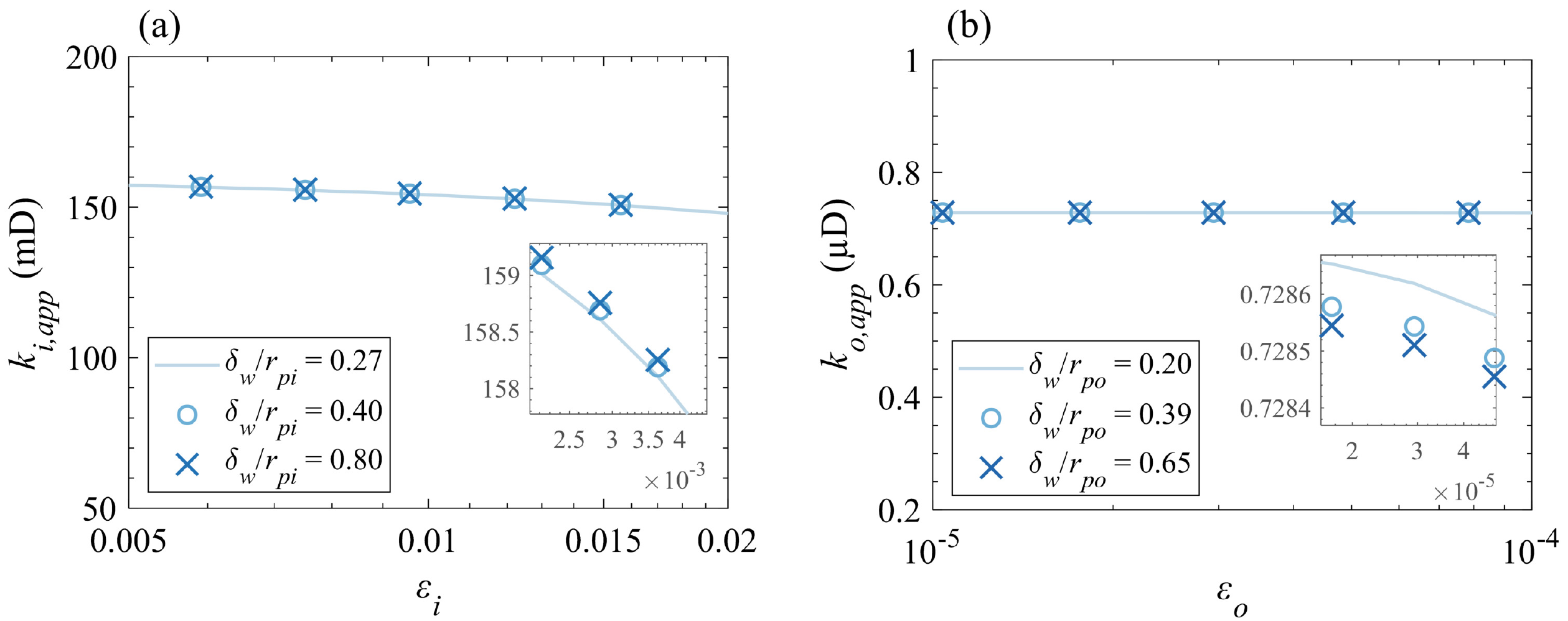}
	\caption{Effect of near-wall thickness in (a) an inorganic pore and (b) an organic pore}
	\label{fig:Sen_nearwall}
\end{figure}

\paragraph{Near-wall thickness.}
Figure \ref{fig:Sen_nearwall} shows the impact of the near-wall regions with respect to pore radii. We observe that a thicker near-wall region in inorganic pores improves the flow capability while in organic pores it reduces the flow capability, although their influences are generally small. This observation is expected because the adhesive interactions between oil and inorganic surfaces are weaker than oil and organic surfaces. 

\begin{figure}[t]
	\centering
	\includegraphics[width=\linewidth]{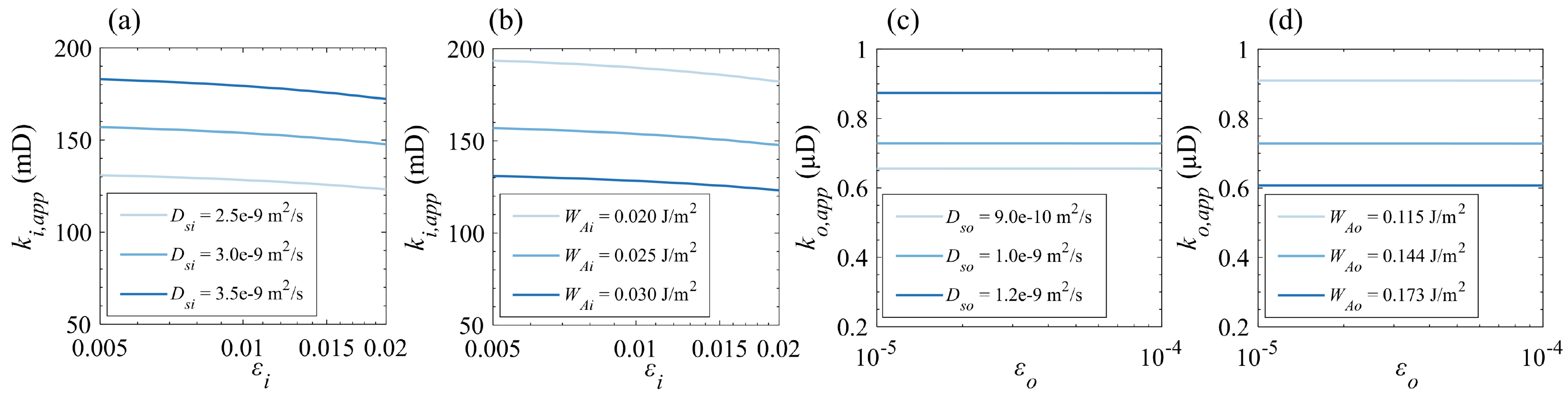}
	\caption{Effect of surface diffusion ($D_s$) and work of adhesion ($W_A$) at increasing roughness ($\varepsilon$). Subscripts $i$ and $o$ denote inorganic and organic pores, respectively. }
	\label{fig:Sen_DsWA}
\end{figure}

\paragraph{Surface diffusion \& work of adhesion.} Figures \ref{fig:Sen_DsWA}(a) and \ref{fig:Sen_DsWA}(c) present the impact of $D_s$ on the ALP in inorganic and organic pores, respectively. The ALP increases with the rise of $D_s$. Oil slippage is more pronounced in inorganic pores (with a higher $D_s$ and a lower $W_A$) than in organic pores of the same diameter, which qualitatively agrees with MD observations of octane's transport through muscovite and kerogen pores \cite{RN24}. Figures \ref{fig:Sen_DsWA}(b) and \ref{fig:Sen_DsWA}(d) show the impact of $W_A$ on the ALP for inorganic and organic pores, respectively. The ALP decreases with the increase of $W_{A}$ due to strong adhesion between liquid and pore surface and the resultant weaker slippage. 

\paragraph{Pore confinement \& surface roughness.}\label{sec:tortuosity} 

\begin{figure}[t]
	\centering
	\includegraphics[width=\linewidth]{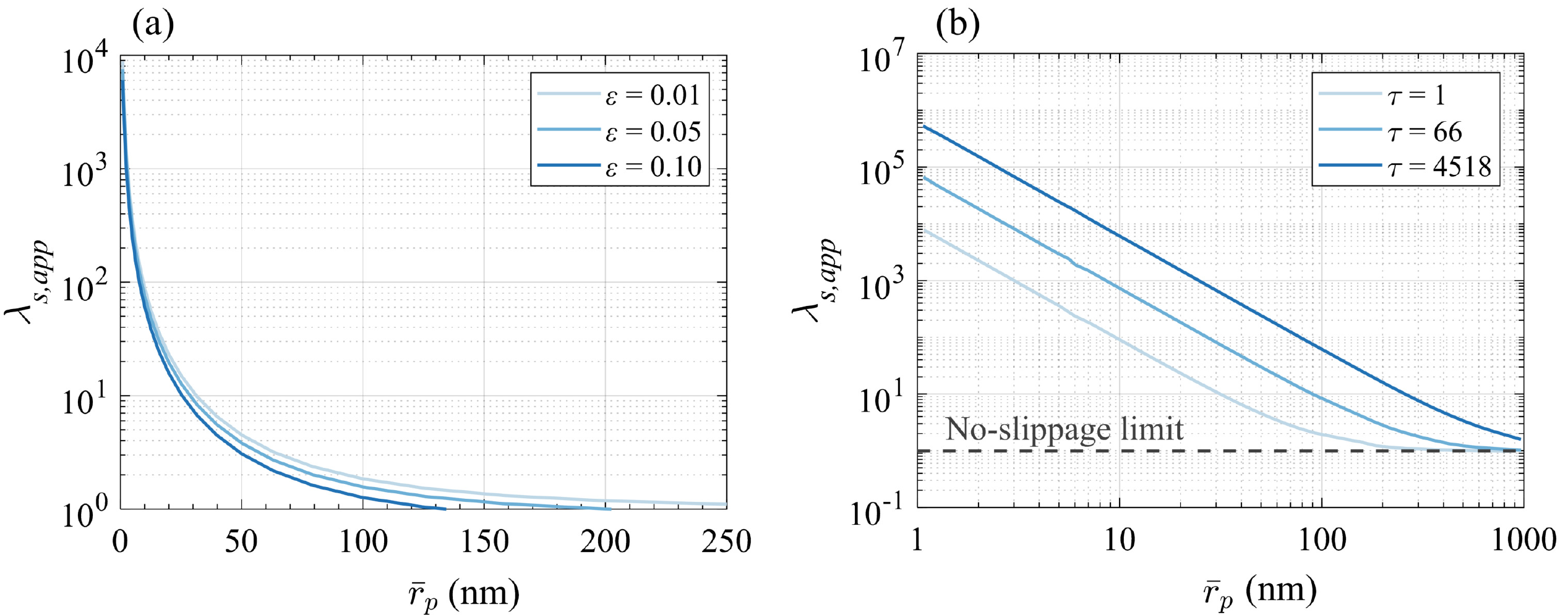}
	\caption{Effect of pore size, tortuosity, and surface roughness on liquid slippage. The calculations are based on $d_{p,max}=10\overline{r}_p$, $\varepsilon =0$, and $\gamma=0.01$.}
	\label{fig:Slippagefactor}
\end{figure}

Figure \ref{fig:Slippagefactor}(a) presents the impact of relative roughness on the apparent slippage factor ($\lambda_{s,app}$). A higher relative roughness leads to a lower $\lambda_{s,app}$, which could quantitatively describe how liquid molecules tend to be ``pinned to'' the irregular wall surface \cite{granick2003slippery}. The sensitivity of ALP to surface roughness depends on pore type since organic pore surfaces are generally smoother and are more uniform than inorganic ones. The ``resistance'' effect of surface roughness is therefore not as evident in smoother organic pores as in rougher inorganic pores (See Figure \ref{fig:Sen_DsWA}).  

The impact of pore confinement on slippage is demonstrated in Figure \ref{fig:Slippagefactor}(b). Slippage is strongly influenced by tortuosity. An increase in tortuosity of $4518/66\approx 68$ or $66/1=66$ can enhance the slippage factor by 10 folds for a pore radius of 1-100 nm. Slippage is also influenced by pore size.  Figure \ref{fig:Slippagefactor}(b) shows that the slippage factor decreases exponentially as the pore radius increases. When the pore radius reaches 100 nm, slippage decreases until no flow enhancement is observed ($\lambda_{s,app} \rightarrow 1$). 

\paragraph{Apparent versus intrinsic liquid permeability.}
\begin{figure}[t]
	\centering
	\includegraphics[width=0.8\linewidth]{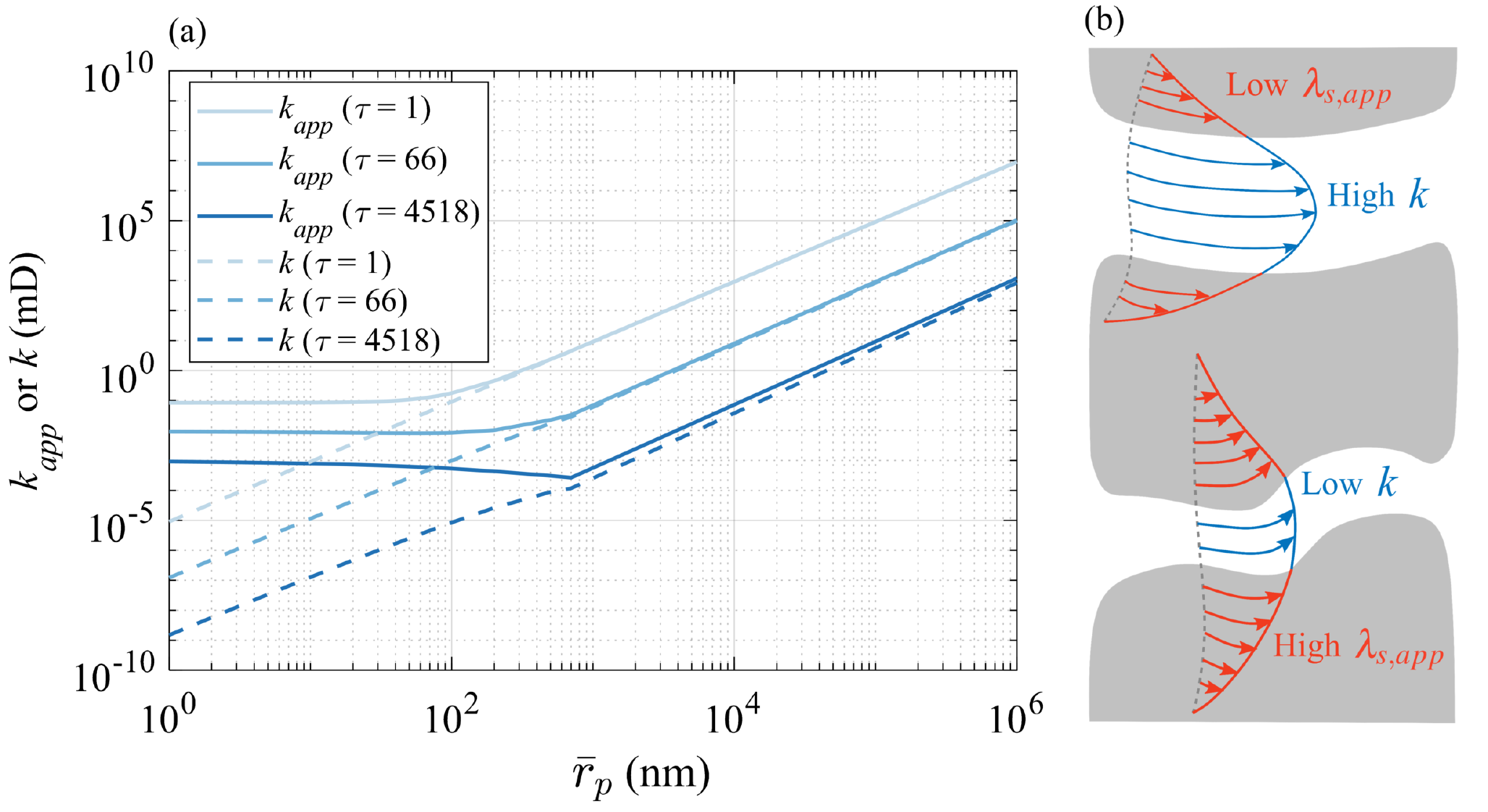}
	\caption{(a) Apparent versus intrinsic liquid permeability under confinement. (b) Pore confinement exerts a negative effect on intrinsic permeability but a positive effect on liquid slippage. Presented are two representative pore-confinement conditions with two pore sizes and two pore throat tortuosities: a weakly confined pore (top) and a strongly confined pore (bottom).}
	\label{fig:AppIntrPerm}
\end{figure}

Figure \ref{fig:AppIntrPerm} shows the overall effect of pore confinement on the apparent and intrinsic liquid permeability, where important observations follow:
\begin{enumerate}
    \item With a decrease in pore throat tortuosity and an increase in pore size, the intrinsic permeability increases.
    \item When the pore size increases, the gap between apparent permeability ($k_{app}$ in solid lines) and intrinsic permeability ($k$ in dashed lines) becomes narrower and lines eventually overlap. This implies that the effect of pore size on flow enhancement fades as the pore size increases. 
    \item When the pore throat tortuosity is increased, the gap between apparent permeability and intrinsic permeability decreases for certain pore sizes, indicating a weakened slippage. 
\end{enumerate}

In Figure \ref{fig:AppIntrPerm}(a), with the increase of $\overline{r}_p$, the points at which the lines of $k$ and $k_{app}$ start to overlap mark the onset of the diminished slippage: $\overline{r}_p\gtrsim100$ nm. This is also the condition of which pore radius  exerts a pronounced positive effect on apparent permeability. We also find that pores with lower intrinsic permeability always have lower apparent permeability, which is because the strong effect of confinement on intrinsic permeability limits the effect of slippage on its apparent permeability even though considerable slippage occurs in highly confined pores. Comparative schematics of the impact of pore confinement on slippage and intrinsic permeability are illustrated in Figure \ref{fig:AppIntrPerm}(b).

\section{Discussion}
\subsection{Liquid transport mechanisms in shales}
Multiple structural and transport factors affect apparent liquid permeability and slippage as indicated by the ALP in Equation \eqref{eq:kapp2}. Dissimilarities in wettability, average pore size, and pore throat tortuosity for pores of different types in mixed-wet porous media further complicate slippage and permeation mechanisms. Table \ref{tab:sum} summarizes the sensitivity comparisons conducted in this work. 

Liquid viscosity near the pore wall can be different from the center of the pore due to the liquid-solid interactions. For example, if the work of adhesion is strong enough where the liquid tends to stick to pore surface, the near-wall viscosity is higher than the viscosity in the pore center. The sensitivity results indicate that viscosity variation near the pore wall may not have a significant impact on the flow enhancement unless the pore diameter is ultra-small, i.e. within an order of liquid molecule size. Given that in shale rocks, the largest connected pores have the most share of the contribution to the overall flow, one may conclude that the fluid viscosity change near the pore wall has a negligible effect compared to surface diffusion and wettability. 

Pore size and its probability distribution as well as pore throat tortuosity are the most dominant structural factors of the ALP. Pore confinement has opposite effects on intrinsic permeability and liquid slippage as it restricts intrinsic permeability but enhances slippage. A quantitative comparison between the estimated range for the intrinsic and apparent permeability suggests that flow enhancement, mostly due to liquid slippage, can reach nearly $300$ in \textit{both} wetting and non-wetting pores. Here, the dual effect of liquid-solid interaction (wettability, adhesion, and surface diffusion) and pore confinement (pore size and pore throat tortuosity) renders such quantitatively comparable flow enhancement in inorganic and organic pores. Nonintuitively, such strong liquid slippage may not necessarily lead to a high apparent permeability when the intrinsic permeability is ultra-low, e.g., of organic matters.

\begin{table}[t]
	\centering
	\caption{Summary of physics for oil slippage and apparent permeability in mixed-wet nanoporous shale}
	\label{tab:sum}\small
\resizebox{\textwidth}{!}{
	\begin{tabular}{llllll}
		\hline
		Parameter   & Physics     & Inorganic pore    & Organic pore     & \multicolumn{2}{l}{Comparison}      \\ \hline
		\multirow{5}{*}{Apparent slippage factor}    & \multirow{3}{*}{Liquid-solid interaction} & Hydrophilic         & Hydrophobic               & \multirow{3}{*}{$\lambda_{si,app} >\lambda_{so,app}$} & \multirow{5}{*}{$\lambda_{si,app} \sim \lambda_{so,app}$} \\ 
		&    & Higher $D_{si}$       & Lower $D_{so}$  & &     \\ 
		&   & Lower $W_{Ai}$   & Higher $W_{Ao}$   &     &      \\ \cline{2-5}
		& \multirow{2}{*}{Pore confinement}         & Higher $\overline{r}_{pi}$ & Lower $\overline{r}_{po}$ & \multirow{2}{*}{$ \lambda_{si,app} < \lambda_{so,app}$} &     \\ 
		&  & Lower $\tau_i$   & Higher $\tau_o$   & &  \\ \hline
		\multirow{2}{*}{Intrinsic permeability} & \multirow{2}{*}{Pore confinement}      & Higher $\overline{r}_{pi}$  & Lower $\overline{r}_{po}$  & \multicolumn{2}{l}{\multirow{2}{*}{$k_{i}\gg k_{o}$}}                                                                                                                  \\ 
		&  & Lower $\tau_i$    & Higher $\tau_o$  & \multicolumn{2}{l}{}                                                                                                                                                   \\ \hline
		Apparent permeability  & \multicolumn{3}{l}{Oil slippage, adsorption, \& intrinsic permeability}   & \multicolumn{2}{l}{$k_{i,app}\gg k_{o,app}$}                                                                                                                           \\ \hline
	\end{tabular}}
\end{table}	 

\subsection{The ALP model comparison with the Carman \& the Carman-Kozeny equation}
It is instructive to understand the relation between the ALP and the fluid equations that predict the pressure drop of fluids through permeable media. We investigate two classic equations, namely Carman \cite{carman1956flow} and Carman-Kozeny \cite{amaefule1993enhanced,carman1937fluid} and show that under reasonable assumptions, the ALP will reduce to the spirit of these two classic equations.

To derive the Carman equation, we begin with a simple version of the ALP. Apply the assumptions of the Carman equation to Equation \eqref{eq:kapp2}, i.e., a constant viscosity distribution ($\mu_b=\mu_w$), no surface diffusion ($D_s=0$), smooth surfaces ($\varepsilon=0$), and uniform pore diameter ($d_p \equiv d_{p,max}$) and set the fractal parameters to $D_T=1$ and $D_p=2$, the ALP reduces to the Carman equation (Equation \eqref{eq:k}) in the limit.

The Carman-Kozeny equation is used for predicting a fluid flowing through permeable media packed with spherical, smooth, and solid grains. A generalized version of the Carman-Kozeny equation is \cite{amaefule1993enhanced}
\begin{equation} \label{eq:kCK}
    k_{C K}=\frac{1}{F_{s} \tau S_{g v}^{2}} \frac{\phi^{3}}{(1-\phi)^{2}}=\frac{d_{p}^{2}}{16 F_{s}} \frac{\phi}{\tau}
\end{equation}
where $k_{CK}$ is the Carman-Kozeny permeability, $F_s$ is the pore shape factor, $S_{gv}$ is the ratio of grain surface area to the grain volume ($S_{g v}=4\phi/d_{p}(1-\phi)$ for spherical grains). Equation \eqref{eq:kCK} accounts for the porosity and geometric properties of grain and pore. The product of $F_s\tau$ is referred to as the Kozeny constant and is a strong function of grain size distribution. The Kozeny constant is often fitted to the experimental data to obtain the best predictor of permeability based on the porosity for different hydraulic units. 
Equation \eqref{eq:k2} has the spirit of the Carman-Kozeny equation in Equation \eqref{eq:kCK} where the fractal function ($\xi$) is the inverse of half the pore shape factor, i.e., $\xi=2/F_s$. For a bundle of identical straight cylinders, i.e., $F_s=2$ and correspondingly $\xi=1$, Equation \eqref{eq:k2} becomes identical to Equation \eqref{eq:kCK}.

\subsection{The ALP versus Klinkenberg equation}
A fundamental insight into the fluid flow in ultra-confined media such as shale rocks, is the presence of fluid slippage, regardless of the phase type. Gas slippage, also known as the Klinkenberg effect, occurs due to the rarefaction. Gas rarefaction is caused by a decrease of gas pressure, reduction of characteristic length, or pore size. Either of these factors increases the dimensionless Knudsen number ($Kn$), defined as a ratio of the mean free path ($\lambda$) of the gas to the average pore diameter ($\overline{d}_p$). With the increase of $Kn$, the gas flow becomes more rarefied and transitions from slip flow to transitional flow, and eventually to the free molecular flow \cite{darabi2012gas,fan2020accurate,javadpour2015gas,fan2017semi}. 

For liquid flows, $\lambda$ is much smaller than $d_p$, $Kn$, therefore, cannot be a proper indicator of liquid slippage. Instead, $W_A$ for liquid has a similar role as $Kn$ for gas. $W_A$ quantifies the energy required to overcome free energies per area of three-phase interfaces of liquid-solid, solid-vapor, and liquid-vapor and subsequently separate liquid phase from the solid phase, $W_A \approx \gamma_{LV}(1+\cos\theta)$ \cite{zisman1964relation}, where $\gamma_{LV}$ is surface tension between the liquid and the saturated vapor in the unit of $N/m$, and $\theta$ is the contact angle between liquid-vapor and liquid-solid interfaces. $\gamma_{LV}$ is related to $\mu_w$, $\theta$ (also wettability), and $D_s$. For example, a large $W_A$ required to separate the liquid from the local site manifests in a small $\theta$, suggesting the liquid is less \textit{willing} to flow near the surface, and slippage is small. The liquid slippage phenomenon is a synergic effect of all the above parameters. 

Based on the sensitivity results, we find that the effect of $\mu_w$ is much smaller than that of $D_s$ and $W_A$. We, therefore, drop the flow contribution from the viscosity. Similarly, $\varepsilon$ can also be dropped as pore-surface roughness is less dominant than the $d_p$ and $\tau$. Given that the fractal intrinsic permeability can essentially represent the Carman-Kozeny equation, the derived ALP in Equation \eqref{eq:kapp2} can be arranged in terms of $k_{CK}$ and $\tau$ as
\begin{equation} \label{eq:kCKapp}
    k_{C K, a p p}=k_{C K}\left(1+\frac{b}{W_{A}}\right),
\end{equation}
where
\begin{equation}
    b=\frac{32 \mu_{b} L_{s} \tau^{\frac{1}{2}}}{\overline{d}_{p}^{2}} D_{s}.
\end{equation}

The simplified ALP model (Equation \eqref{eq:kCKapp}) for liquid presents some interesting analogies as the Klinkenberg equation for gas \cite{klinkenberg1941permeability}. First, the liquid slippage is inversely proportional to $W_A$ whereas the gas slippage is inversely proportional to the gas pressure. Second, the term $b$, defined here as the liquid slippage constant, determines the flow enhancement contribution upon the intrinsic $k_{CK}$. It is a function of pore confinement, i.e., $\overline{d}_p$ and $\tau$, and liquid $D_s$. Interestingly, the term $b$ is similar to the gas slippage constant ($b'$) in the Klinkenberg equation as $b'$ is found to be a strong function of $\tau$ and the gas-solid interaction parameter -- the tangential momentum accommodation coefficient (TMAC) \cite{wu2017apparent}, where the TMAC characterizes the how gas molecules are reflected in terms of diffuse reflection and specular reflection on the wall after the gas-wall collision \cite{arkilic2001mass}. 

In a more general manner, by assuming that liquid would follow the hydraulic pathways of the pores, we define a dimensionless parameter, the liquid confinement number ($Cn$), as the ratio of the tortuous path length to the characteristic length, to characterize the liquid slippage. In practice, the average pore diameter is applied as the characteristic length, therefore $C n=L_{s} \sqrt{\tau} / \overline{d}_{p}$. Equation \eqref{eq:kCKapp} then reads 
\begin{equation} \label{eq:kCKapp2}
    k_{C K, a p p}=k_{C K}(1+\alpha' \cdot C n)
\end{equation}
where the dimensionless parameter 
\begin{equation}
    \alpha'=\frac{32 \mu_{b} D_{s}}{\overline{d}_{p} W_{A}}
\end{equation}
quantifies the surface diffusion of liquid of viscosity $\mu_b$ in the straight pore of the diameter $\overline{d}_{p}$ by overcoming the work of adhesion $W_A$. Equation \eqref{eq:kCKapp2} shares a similar structure as the gas slippage model due to gas rarefaction \cite{beskok1999report}. 

The derived ALP model in Equation \eqref{eq:kapp2} along with its transformation in Equations \eqref{eq:kCKapp} and \eqref{eq:kCKapp2} delivers a more comprehensive description of the liquid flow in tortuous, heterogeneous porous media, and under proper restricting assumptions, reduces to the spirit of the Carman-Kozeny equation.

\section{Conclusions}
Liquid transport in shale rocks is governed by local pore confinement, liquid-solid interaction, and pore-surface roughness. We proposed an apparent liquid permeability (ALP) model for heterogeneous and rough nanoporous shale matrices, and a workflow for the ALP estimation. Major conclusions follow:
\begin{enumerate}
    \item Inorganic pores and organic pores require separate modeling as they possess different pore size distribution, pore throat tortuosity, pore-surface roughness, pore surface wettability, and liquid-solid interaction.
    \item Liquid slippage on a wetting surface is enhanced for a high pore-confinement effect, e.g., the strength of oil slippage in organic pores is quantitatively considerable to that in inorganic pores, due to high pore confinements.
    \item Apparent permeability is restricted by high pore confinements.
    \item Oil slippage abates when pore-surface roughness intensifies. 
    \item The ALP model shares some analogies with the Klinkenberg gas permeability and also converges to the Carman-Kozeny permeability when no-slip liquid flows through a bundle of homogeneous capillaries.
\end{enumerate}

\clearpage
\section*{Nomenclature}
\begin{table}[h]
\begin{tabular}{llllllllllll}
\hline
$d_p$         & Pore diameter                                              \\
$r_p$         & Pore radius                                                \\
$\delta_w$    & Near-wall region thickness                                 \\
$L_s$         & Straight pore length                                       \\
$L_p$         & Tortuous pore length                                       \\
$\tau$        & Tortuosity                                                 \\
$\phi$        & Porosity                                                   \\
$d_m$         & Matrix grain diameter                                      \\
$\alpha$      & Areal ratio of the conical nanostructures (roughness elements)                     \\
$\beta$       & Ratio of the minimum to the maximum conical base diameter  \\
$\gamma$      & Pore-size heterogeneity coefficient                        \\
$\varepsilon$ & Relative roughness                                         \\
$\alpha_s$    & Fraction of the available sites for liquid migration       \\
$\alpha^*$    & Bending angle of the tube                                  \\
$\beta^*$     & Tilting angle of the tube                                  \\
$D_p$         & Pore size fractal dimension                                \\
$D_T$         & Tortuosity fractal dimension                                  \\
$D_c$         & Fractal dimension of conical base size distribution        \\
$\mu_b$       & Bulk viscosity                                             \\
$\mu_w$       & Near-wall viscosity                                        \\ 
$D_s$         & Surface diffusion coefficient  \\
$W_A$         & Work of adhesion                \\
$\Delta P$    & Pressure difference             \\
$Q$           & Volumetric flow rate            \\
$\lambda_b$ & Pore-structure factor\\
$\lambda_s$ & Slippage factor\\
$l_{slip}$ & Slip length\\ \\
Subscript     &                                 \\
$app$           & Apparent                        \\
$i$             & Inorganic matter                \\
$o$             & Organic matter                  \\\hline
\end{tabular}
\end{table}
\clearpage

\section*{Appendix}
\appendix
\section{Review of pore-scale fractal models and the intrinsic permeability model} \label{sec:fractal}
\setcounter{equation}{0}
\renewcommand{\theequation}{A.\arabic{equation}}

In an REV, pore space is conventionally modeled as a bundle of cylindrical tubes with a constant diameter ($d_p$) and a length ($L_p$). For a laminar viscous flow through an REV, a no-slip boundary condition applies to solve the flow rate as
\begin{equation}\label{eq:q}
		Q = \frac{\pi d_p^4\Delta P}{128\mu_b L_p},
	\end{equation}
where $Q$ is the intrinsic volumetric flow rate; $d_p$ is pore diameter; $L_p$ is tortuous pore length; $\Delta P$ is pressure difference; $\mu_b$ is fluid viscosity. By applying Darcy's law to Equation \eqref{eq:q}, permeability is solved as Carman's permeability \cite{carman1956flow}:
	\begin{equation}\label{eq:k}
		k = \frac{\pi d_p^4}{128A},
	\end{equation}
where $A$ is the cross-sectional area of the REV.
		
The fractal theory is applied to model REV heterogeneities in pore throat tortuosity, pore size distribution, and roughness of pore surface \cite{yang2015permeability,yu2002fractal}. First, pore throat tortuosity is modeled by a tortuosity fractal dimension ($D_T$) that relates the straight pore length ($L_s$) to tortuous pore length ($L_p$) as shown in Equation \eqref{eq:Lp}. Second, the cumulative number of pores with a diameter greater than $d_p$ is modeled as a function of the ratio of the maximum pore diameter ($d_{p,max}$) to the variable $d_p$ in the REV, to the power of the pore size fractal dimension ($D_p$), as shown in Equation \eqref{eq:N_p}. Third, we model pore-surface roughness as numerous conical nanostructures protruding from the inner surface of spherical pores, in which the cumulative number ($N_c$) of such nanostructures is a function of the ratio of the maximum conical base diameter ($d_{c,max}$) to the variable ($d_{c}$) in the local pore, to the power of the fractal dimension of the conical base size distribution ($D_c$), shown in Equation \eqref{eq:N_c}.  	
	\begin{equation}\label{eq:Lp}
		L_p(d_p) = d_p^{-D_T+1}L_s^{D_T}
	\end{equation}
	\begin{equation}\label{eq:N_p}
		N_p(d\geq d_p) = (\frac{d_{p,max}}{d_p})^{D_p}
	\end{equation} 	\begin{equation}\label{eq:N_c}
		N_c(d\geq d_c) = (\frac{d_{c,max}}{d_c})^{D_c}
	\end{equation} 
	
By combining Equations \eqref{eq:k} through \eqref{eq:N_c}, intrinsic permeability is derived as \cite{yang2015permeability}:
\begin{equation}\label{eq:k1}
	k= \frac{\pi d_{p,max}^{D_{T}+3}D_{p}L_{s}^{-D_{T}+1}(1-\varepsilon)^4}{128A(D_{T}-D_{p}+3)},
\end{equation}
where $\varepsilon$ is the relative roughness, defined as the ratio of double of the average height of conical nanostructures in a pore to its pore diameter $d_p$, i.e., $\varepsilon = 2(\overline{h}_c)_{d_p}/d_{p}$. By solving for $(\overline{h}_c)_{d_p}$, $\varepsilon$ is derived as Equation \eqref{eq:e}. Derivation of $\varepsilon$ is referred to \cite{yang2015permeability,yang2014fractal}.

\begin{equation}\label{eq:e}
\varepsilon = \frac{2\alpha }{3}\frac{(h_{c,max})_{d_{p,min}}}{d_{p,min}}\frac{2-D_c}{3-D_c}\frac{1-\beta^{-D_c+3}}{1-\beta^{-D_c+2}},
\end{equation}
where $(h_{c,max})_{d_{p,min}}$ is the maximum height of the cone in the minimum pore diameter ($d_{p,min}$); $\alpha$ is the ratio of the total cone base area ($S_{c1}+S_{c2}+S_{c3}+S_{c4}+...$) to the total pore surface area ($S_p$) (including protruding cone base area and non-protruding smooth area) in Figure \ref{fig:Workflow}(c); and $\beta$ is the ratio of the minimum to the maximum base diameter, given by $\beta = d_{c,min}/d_{c,max}$. With $D_c$ approaching 0, fewer conical nanostructures occupy pore surface, therefore pore-surface roughness decreases. 

In Equation \eqref{eq:k1}, the cross-sectional area of an REV ($A$)  cannot be measured directly: a common approach is to substitute $A$ with porosity $\phi$ as follows. Considering $N_p$ numbers of tortuous cylinders for 3D pores in an REV, we calculate porosity by the volumetric ratio of pore space over the REV:
\begin{equation}\label{eq:phi}
	\begin{aligned}
		\phi &= \frac{-\int_{d_{p,min}}^{d_{p,max}} \Big[\pi d_{p}^2L_{p} (d_{p})\Big] \mathbf{d}N_{p}(d_{p})}{4AL_{s}} \\
		&= \frac{\pi d_{p,max}^{-D_T+3} D_p L_s^{D_T-1} (1-\gamma^{-D_T-D_p+3})}{4A(-D_T-D_p+3)},
	\end{aligned}
\end{equation}
where $\gamma$ is the pore-size heterogeneity coefficient, defined as the ratio of the minimum to the maximum pore diameter in an REV, i.e., $\gamma = d_{p,min}/d_{p,max}$. By Equation \eqref{eq:phi}, $A$ is derived:
\begin{equation}\label{eq:A}
	A = \frac{\pi d_{p,max}^{-D_T+3} D_p L_s^{D_T-1} (1-\gamma^{-D_T-D_p+3})}{4\phi(-D_T-D_p+3)}.
\end{equation}
Substituting $A$ in Equation \eqref{eq:k1} with \eqref{eq:A}, one can derive the intrinsic permeability as in Equation \eqref{eq:k2}.

\section{Derivation of flow enhancement} \label{sec:slip}
\setcounter{equation}{0}
\renewcommand{\theequation}{B.\arabic{equation}}
The apparent volumetric flow rate in a pore is solved as \cite{zhang2017apparent} 
\begin{equation}\label{eq:Qapp}
	\begin{aligned}
		Q_{app} &= \frac{\pi \Delta P}{8L_{p}}\Big\{\frac{(r_{p}-\delta_w)^2}{\mu_b}\Big[\frac{\mu_b}{\mu_w}(4r_{p}\delta_w-2\delta_w^2)+(r_{p}-\delta_w)^2+\frac{8\mu_bD_sL_{p}}{W_A}\Big]\\
		& + \frac{1}{\mu_w}(2r_{p}\delta_w-\delta_w^2)(2r_{p}\delta_w-\delta_w^2+\frac{8\mu_wD_sL_{p}}{W_A})\Big\},
	\end{aligned} 
\end{equation}
where $D_s$ is the surface diffusion coefficient. A high value of $D_{s}$ reflects a fast diffusion of liquid molecules on the surface. Measured $D_{s}$ values for oil on different wettability surfaces are reported in the order of 1$\times10^{-9}$ m$^2$/s to 1$\times10^{-8}$ m$^2$/s \cite{mattia2012explaining,ershov2001displacement}.

\setcounter{table}{0}
\renewcommand{\thetable}{C.\arabic{table}}
\begin{sidewaystable}
    \caption{Literature data (MD and AFM) for hydrocarbon liquid physical and transport properties on hydrophilic and hydrophobic surfaces. }\label{tab:MDdatasummary}
\begin{threeparttable}
\begin{tabular}{p{1.5cm}p{1.5cm}p{1.2cm}p{1cm}p{1cm}p{0.7cm}p{1cm}p{1cm}p{0.8cm}p{1cm}p{1cm}p{1.5cm}}
\hline
Wettability & Substrate   & Fluid  & $T$(K) & $P$ (MPa) & $H$ (nm) & $W_A$ (mJ/m$^2$) & $D_s$ (10$^{-9}$ m$^2$/s) & $l_{slip}$ (nm) & 2$\delta_w/H$ (nm/nm) & Data type & Reference    \\ \hline
Hydrophilic & Muscovite   & Octane & 300   & 20   & $\sim$3.58   & - & -  & 0.6 & 0.82  & MD  & \cite{RN24}   \\
Hydrophilic & Silica   & Octane & 298   & - & 2  & - & -  & -   & -  & MD  & \cite{RN25}   \\
Hydrophilic & Silica   & Octane & 353   & 30   & 5.24  & - & 2.88  & 0.9 & 0.48  & MD  &\cite{RN23}  \\
Hydrophilic & Silica   & Octane & 383   & 30   & 2  & - & -  & -   & -  & MD  & \cite{RN26} \\
Hydrophilic & Silica   & Octane & 383   & 30   & 5  & - & -  & -   & 0.49  & MD  & \cite{RN26}  \\
Hydrophilic & Silica   & Octane & 383   & 30   & 10 & - & -  & -   & 0.28  & MD  & \cite{RN26}  \\
Hydrophilic & Silica   & Octane & 383   & 30   & 20 & - & -  & -   & 0.12  & MD  & \cite{RN26}  \\
Hydrophilic & Silica   & Hexane\tnote{a}   & 400   & - & -  & - & 2.3   & -   & -  & MD  & \cite{RN27} \\
Hydrophilic & Silica   & Hexane\tnote{b}   & 400   & - & -  & - & 2.7   & -   & -  & MD  & \cite{RN27} \\
Hydrophilic & Silica   & Hexane\tnote{c}   & 400   & - & -  & - & 3.3   & -   & -  & MD  & \cite{RN27} \\
Hydrophilic & Silica   & Hexane\tnote{d}   & 400   & - & -  & - & 3.8   & -   & -  & MD  & \cite{RN27} \\
Hydrophilic & Calcite  & Hexadecyl & -  & - & -  & 0 & -  & -   & -  & AFM & \cite{RN28} \\
Hydrophilic & Inorganic Chalk    & Hexadecyl & -  & - & -  & 3 & -  & -   & -  & AFM & \cite{RN28} \\ \hline
Hydrophobic & Kerogen (activated)   & Octane & 300   & 20   & \textless{}3.9  & - & -  & 0   & -  & MD  & \cite{RN24}   \\
Hydrophobic & Kerogen (activated)   & Octane & 300   & 30   & 5  & - & -  & -   & -  & MD  & \cite{RN29} \\
Hydrophobic & Kerogen (inactivated) & Octane & 300   & 30   & 5  & - & -  & -   & 0.40  & MD  & \cite{RN29}  \\
Hydrophobic & Graphene & Octane & 353   & 30   & 5.24  & - & -  & 132.48 & 0.38  & MD  & \cite{RN23}   \\
Hydrophobic & Graphite & Octane & 298   & - & 2  & - & -  & -   & 0.40  & MD  & \cite{RN25}   \\
Hydrophobic & Graphite & Octane & 300   & 30   & 5  & - & -  & -   & 0.52  & MD  & \cite{RN29}  \\
Hydrophobic & Organic Chalk   & Hexadecyl & -  & - & -  & 115   & -  & -   & -  & AFM & \cite{RN28} \\ \hline
      \end{tabular}
      \begin{tablenotes}
        \footnotesize
        \item Nomenclature: $T$ = temperature; $P$ = Pressure; $H$ = slit aperture.  
        \item[a] in a 47\% hexane-53\% hexadecane molar fraction mixture.
        \item[b]  in a 64\% hexane-36\% hexadecane molar fraction mixture.
        \item[c] in an 85\% hexane-15\% hexadecane molar fraction mixture.
        \item[d] in a 47\% hexane-53\% octane molar fraction mixture.
      \end{tablenotes}
    \end{threeparttable}
  \end{sidewaystable}

\clearpage
\bibliography{bibli}   

\begin{thebibliography}{10}

\bibitem{de2002fluid}
Pierre-Gilles de~Gennes.
\newblock On fluid/wall slippage.
\newblock {\em Langmuir}, 18(9):3413--3414, 2002.

\bibitem{joseph2008carbon}
Sony Joseph and NR~Aluru.
\newblock Why are carbon nanotubes fast transporters of water?
\newblock {\em Nano letters}, 8(2):452--458, 2008.

\bibitem{myers2011slip}
Tim~G Myers.
\newblock Why are slip lengths so large in carbon nanotubes?
\newblock {\em Microfluidics and nanofluidics}, 10(5):1141--1145, 2011.

\bibitem{podolska2013water}
NI~Podolska and AI~Zhmakin.
\newblock Water flow in micro-and nanochannels. molecular dynamics simulations.
\newblock In {\em Journal of Physics: Conference Series}, volume 461. IOP
  Publishing, 2013.
\newblock 012034.

\bibitem{whitby2007fluid}
M~Whitby and N~Quirke.
\newblock Fluid flow in carbon nanotubes and nanopipes.
\newblock {\em Nature Nanotechnology}, 2(2):87--94, 2007.

\bibitem{hummer2001water}
Gerhard Hummer, Jayendran~C Rasaiah, and Jerzy~P Noworyta.
\newblock Water conduction through the hydrophobic channel of a carbon
  nanotube.
\newblock {\em Nature}, 414(6860):188, 2001.

\bibitem{majumder2005nanoscale}
Mainak Majumder, Nitin Chopra, Rodney Andrews, and Bruce~J Hinds.
\newblock Nanoscale hydrodynamics: enhanced flow in carbon nanotubes.
\newblock {\em Nature}, 438(7064):44--44, 2005.

\bibitem{striolo2006mechanism}
Alberto Striolo.
\newblock The mechanism of water diffusion in narrow carbon nanotubes.
\newblock {\em Nano Letters}, 6(4):633--639, 2006.

\bibitem{falk2010molecular}
Kerstin Falk, Felix Sedlmeier, Laurent Joly, Roland~R Netz, and Lyd{\'e}ric
  Bocquet.
\newblock Molecular origin of fast water transport in carbon nanotube
  membranes: superlubricity versus curvature dependent friction.
\newblock {\em Nano letters}, 10(10):4067--4073, 2010.

\bibitem{noy2007nanofluidics}
Aleksandr Noy, Hyung~Gyu Park, Francesco Fornasiero, Jason~K Holt, Costas~P
  Grigoropoulos, and Olgica Bakajin.
\newblock Nanofluidics in carbon nanotubes.
\newblock {\em Nano today}, 2(6):22--29, 2007.

\bibitem{ho2011liquid}
Tuan~Anh Ho, Dimitrios~V Papavassiliou, Lloyd~L Lee, and Alberto Striolo.
\newblock Liquid water can slip on a hydrophilic surface.
\newblock {\em Proceedings of the National Academy of Sciences}, 108(39), 2011.
\newblock 16170--16175.

\bibitem{blake1990slip}
Terence~D Blake.
\newblock Slip between a liquid and a solid: {DM T}olstoi's (1952) theory
  reconsidered.
\newblock {\em Colloids and surfaces}, 47:135--145, 1990.

\bibitem{thomas2008reassessing}
John~A Thomas and Alan~JH McGaughey.
\newblock Reassessing fast water transport through carbon nanotubes.
\newblock {\em Nano letters}, 8(9):2788--2793, 2008.

\bibitem{mattia2012explaining}
Davide Mattia and Francesco Calabr{\`o}.
\newblock Explaining high flow rate of water in carbon nanotubes via
  solid--liquid molecular interactions.
\newblock {\em Microfluidics and nanofluidics}, 13(1):125--130, 2012.

\bibitem{fan2016transient}
Dian Fan and Amin Ettehadtavakkol.
\newblock Transient shale gas flow model.
\newblock {\em Journal of Natural Gas Science and Engineering}, 33:1353--1363,
  2016.

\bibitem{fan2017analytical}
Dian Fan and Amin Ettehadtavakkol.
\newblock Analytical model of gas transport in heterogeneous
  hydraulically-fractured organic-rich shale media.
\newblock {\em Fuel}, 207:625--640, 2017.

\bibitem{rezaee2015fundamentals}
Reza Rezaee.
\newblock {\em Fundamentals of gas shale reservoirs}.
\newblock John Wiley \& Sons, 2015.

\bibitem{song2019nonlinear}
Fuquan Song, Liwen Bo, Shiming Zhang, and Yeheng Sun.
\newblock Nonlinear flow in low permeability reservoirs: Modelling and
  experimental verification.
\newblock {\em Advances in Geo-Energy Research}, 3(1):76--81, 2019.

\bibitem{christensen2017enhanced}
M~Christensen and Y~Tanino.
\newblock Enhanced permeability due to apparent oil/brine slippage in limestone
  and its dependence on wettability.
\newblock {\em Geophysical Research Letters}, 44(12):6116--6123, 2017.

\bibitem{cui2019oil}
Jiangfeng Cui.
\newblock Oil transport in shale nanopores and micro-fractures: Modeling and
  analysis.
\newblock {\em Journal of Petroleum Science and Engineering}, 178:640--648,
  2019.

\bibitem{cui2017liquid}
Jiangfeng Cui, Qian Sang, Yajun Li, Congbin Yin, Yanchao Li, and Mingzhe Dong.
\newblock Liquid permeability of organic nanopores in shale: Calculation and
  analysis.
\newblock {\em Fuel}, 202:426--434, 2017.

\bibitem{fan2019enhanced}
Haiming Fan, Hui Li, and Han Wang.
\newblock Enhanced oil flow model coupling fractal roughness and heterogeneous
  wettability.
\newblock {\em Fractals}, 2019.

\bibitem{feng2019apparent}
Qihong Feng, Shiqian Xu, Sen Wang, Yuyao Li, Fangfang Gao, and Yajuan Xu.
\newblock Apparent permeability model for shale oil with multiple mechanisms.
\newblock {\em Journal of Petroleum Science and Engineering}, 175:814--827,
  2019.

\bibitem{wang2019apparent}
Han Wang, Yuliang Su, Zhenfeng Zhao, Wendong Wang, Guanglong Sheng, and Shiyuan
  Zhan.
\newblock Apparent permeability model for shale oil transport through elliptic
  nanopores considering wall-oil interaction.
\newblock {\em Journal of Petroleum Science and Engineering}, 176:1041--1052,
  2019.

\bibitem{wang2019fractal}
Qing Wang and Zhilin Cheng.
\newblock A fractal model of water transport in shale reservoirs.
\newblock {\em Chemical Engineering Science}, 198:62--73, 2019.

\bibitem{yang2019pore}
Yongfei Yang, Ke~Wang, Lei Zhang, Hai Sun, Kai Zhang, and Jingsheng Ma.
\newblock Pore-scale simulation of shale oil flow based on pore network model.
\newblock {\em Fuel}, 251:683--692, 2019.

\bibitem{zhang2017apparent}
Qi~Zhang, Yuliang Su, Wendong Wang, Mingjing Lu, and Guanglong Sheng.
\newblock Apparent permeability for liquid transport in nanopores of shale
  reservoirs: Coupling flow enhancement and near wall flow.
\newblock {\em International Journal of Heat and Mass Transfer}, 115:224--234,
  2017.

\bibitem{lu2015organic}
Jiemin Lu, Stephen~C Ruppel, and Harry~D Rowe.
\newblock Organic matter pores and oil generation in the {Tuscaloosa} marine
  shale.
\newblock {\em AAPG Bulletin}, 99(2):333--357, 2015.

\bibitem{chalmers2012characterization}
Gareth~R Chalmers, R~Marc Bustin, and Ian~M Power.
\newblock Characterization of gas shale pore systems by porosimetry,
  pycnometry, surface area, and field emission scanning electron
  microscopy/transmission electron microscopy image analyses: {Examples} from
  the {Barnett, Woodford, Haynesville, Marcellus, and Doig} units.
\newblock {\em AAPG bulletin}, 96(6):1099--1119, 2012.

\bibitem{javadpour2015slip}
F~Javadpour, M~McClure, and ME~Naraghi.
\newblock Slip-corrected liquid permeability and its effect on hydraulic
  fracturing and fluid loss in shale.
\newblock {\em Fuel}, 160:549--559, 2015.

\bibitem{cao2006liquid}
Bing-Yang Cao, Min Chen, and Zeng-Yuan Guo.
\newblock Liquid flow in surface-nanostructured channels studied by molecular
  dynamics simulation.
\newblock {\em Physical Review E}, 74(6), 2006.
\newblock 066311.

\bibitem{ambrose2010new}
Raymond~Joseph Ambrose, Robert~Chad Hartman, Mery Diaz~Campos, I~Yucel Akkutlu,
  and Carl Sondergeld.
\newblock New pore-scale considerations for shale gas in place calculations.
\newblock In {\em SPE Unconventional Gas Conference}. Society of Petroleum
  Engineers, 2010.

\bibitem{wang2016molecular}
Sen Wang, Farzam Javadpour, and Qihong Feng.
\newblock Molecular dynamics simulations of oil transport through inorganic
  nanopores in shale.
\newblock {\em Fuel}, 171:74--86, 2016.

\bibitem{RN14}
Amir~Barati Farimani, Mohammad Heiranian, and Narayana~R Aluru.
\newblock Nano-electro-mechanical pump: Giant pumping of water in carbon
  nanotubes.
\newblock {\em Scientific reports}, 6:26211, 2016.

\bibitem{RN22}
M~Shaat.
\newblock Viscosity of water interfaces with hydrophobic nanopores: application
  to water flow in carbon nanotubes.
\newblock {\em Langmuir}, 33(44):12814--12819, 2017.

\bibitem{RN17}
Dhaval~A Doshi, Erik~B Watkins, Jacob~N Israelachvili, and Jaroslaw Majewski.
\newblock Reduced water density at hydrophobic surfaces: Effect of dissolved
  gases.
\newblock {\em Proceedings of the National Academy of Sciences},
  102(27):9458--9462, 2005.

\bibitem{RN15}
Ji{\v{r}}{\'i} Jane{\v{c}}ek and Roland~R Netz.
\newblock Interfacial water at hydrophobic and hydrophilic surfaces: Depletion
  versus adsorption.
\newblock {\em Langmuir}, 23(16):8417--8429, 2007.

\bibitem{RN19}
Torben~R Jensen, Morten~{\o}stergaard Jensen, Niels Reitzel, Konstantin
  Balashev, G{\"{u}}nther~H Peters, Kristian Kjaer, and Thomas Bj{\o}rnholm.
\newblock Water in contact with extended hydrophobic surfaces: direct evidence
  of weak dewetting.
\newblock {\em Physical Review Letters}, 90(8):086101, 2003.

\bibitem{RN18}
Shavkat~I Mamatkulov, Pulat~K Khabibullaev, and Roland~R Netz.
\newblock Water at hydrophobic substrates: curvature, pressure, and temperature
  effects.
\newblock {\em Langmuir}, 20(11):4756--4763, 2004.

\bibitem{RN16}
Christian Sendner, Dominik Horinek, Lyderic Bocquet, and Roland~R Netz.
\newblock Interfacial water at hydrophobic and hydrophilic surfaces: Slip,
  viscosity, and diffusion.
\newblock {\em Langmuir}, 25(18):10768--10781, 2009.

\bibitem{RN23}
Sen Wang, Farzam Javadpour, and Qihong Feng.
\newblock Molecular dynamics simulations of oil transport through inorganic
  nanopores in shale.
\newblock {\em Fuel}, 171:74--86, 2016.

\bibitem{ghanbarian2013tortuosity}
Behzad Ghanbarian, Allen~G Hunt, Robert~P Ewing, and Muhammad Sahimi.
\newblock Tortuosity in porous media: a critical review.
\newblock {\em Soil science society of America journal}, 77(5):1461--1477,
  2013.

\bibitem{Wei2015An}
Wei Wei, Jianchao Cai, Xiangyun Hu, and Qi~Han.
\newblock An electrical conductivity model for fractal porous media.
\newblock {\em Geophysical Research Letters}, 42(12):4833--4840, 2015.

\bibitem{yu2008analysis}
Boming Yu.
\newblock Analysis of flow in fractal porous media.
\newblock {\em Applied Mechanics Reviews}, 61(5), 2008.
\newblock 050801.

\bibitem{fan2019confinement}
Dian Fan, Wendong Wang, Amin Ettehadtavakkol, and Yuliang Su.
\newblock Confinement facilitates wetting liquid slippage through mixed-wet and
  heterogeneous nanoporous shale rocks.
\newblock In {\em Unconventional Resources Technology Conference (URTEC)}.
  Society of Petroleum Engineers. American Association of Petroleum Geologists.
  Society of Exploration Geophysicists., 2019.

\bibitem{fan2018fluids}
Dian Fan.
\newblock {\em Fluids Transport in Heterogeneous Shale Rocks}.
\newblock PhD thesis, Texas Tech University Lubbock, TX, 2018.

\bibitem{chen2015nanoscale}
Li~Chen, Lei Zhang, Qinjun Kang, Hari~S Viswanathan, Jun Yao, and Wenquan Tao.
\newblock Nanoscale simulation of shale transport properties using the lattice
  {Boltzmann} method: permeability and diffusivity.
\newblock {\em Scientific Reports}, 5:8089, 2015.

\bibitem{yang2015new}
Yongfei Yang, Jun Yao, Chenchen Wang, Ying Gao, Qi~Zhang, Senyou An, and Wenhui
  Song.
\newblock New pore space characterization method of shale matrix formation by
  considering organic and inorganic pores.
\newblock {\em Journal of Natural Gas Science and Engineering}, 27:496--503,
  2015.

\bibitem{josh2012laboratory}
M~Josh, L~Esteban, C~Delle~Piane, J~Sarout, DN~Dewhurst, and MB~Clennell.
\newblock Laboratory characterisation of shale properties.
\newblock {\em Journal of Petroleum Science and Engineering}, 88:107--124,
  2012.

\bibitem{yang2014fractal}
Shanshan Yang, Boming Yu, Mingqing Zou, and Mingchao Liang.
\newblock A fractal analysis of laminar flow resistance in roughened
  microchannels.
\newblock {\em International Journal of Heat and Mass Transfer}, 77:208--217,
  2014.

\bibitem{yang2015permeability}
Shanshan Yang, Mingchao Liang, Boming Yu, and Mingqing Zou.
\newblock Permeability model for fractal porous media with rough surfaces.
\newblock {\em Microfluidics and Nanofluidics}, 18(5-6):1085--1093, 2015.

\bibitem{chen2015permeability}
Li~Chen, Qinjun Kang, Zhenxue Dai, Hari~S Viswanathan, and Wenquan Tao.
\newblock Permeability prediction of shale matrix reconstructed using the
  elementary building block model.
\newblock {\em Fuel}, 160:346--356, 2015.

\bibitem{whitby2008enhanced}
Max Whitby, Laurent Cagnon, Maya Thanou, and Nick Quirke.
\newblock Enhanced fluid flow through nanoscale carbon pipes.
\newblock {\em Nano letters}, 8(9):2632--2637, 2008.

\bibitem{hassenkam2009probing}
Tue Hassenkam, Lone~Lindb{\ae}k Skovbjerg, and Susan Louise~Svane Stipp.
\newblock Probing the intrinsically oil-wet surfaces of pores in north sea
  chalk at subpore resolution.
\newblock {\em Proceedings of the National Academy of Sciences},
  106(15):6071--6076, 2009.

\bibitem{chen2015pore}
Li~Chen, Qinjun Kang, Rajesh Pawar, Ya-Ling He, and Wen-Quan Tao.
\newblock Pore-scale prediction of transport properties in reconstructed
  nanostructures of organic matter in shales.
\newblock {\em Fuel}, 158:650--658, 2015.

\bibitem{wang2015oil}
Sen Wang, Qihong Feng, Farzam Javadpour, Tian Xia, and Zhen Li.
\newblock Oil adsorption in shale nanopores and its effect on recoverable
  oil-in-place.
\newblock {\em International Journal of Coal Geology}, 147:9--24, 2015.

\bibitem{ershov2001displacement}
AP~Ershov, ZM~Zorin, VD~Sobolev, and NV~Churaev.
\newblock Displacement of silicone oils from the hydrophobic surface by aqueous
  trisiloxane solutions.
\newblock {\em Colloid Journal}, 63(3):290--295, 2001.

\bibitem{naraghi2015stochastic}
Morteza~Elahi Naraghi and Farzam Javadpour.
\newblock A stochastic permeability model for the shale-gas systems.
\newblock {\em International Journal of Coal Geology}, 140:111--124, 2015.

\bibitem{shen2007critical}
Lihua Shen and Zhangxin Chen.
\newblock Critical review of the impact of tortuosity on diffusion.
\newblock {\em Chemical Engineering Science}, 62(14):3748--3755, 2007.

\bibitem{RN1}
Nils~R Backeberg, Francesco Iacoviello, Martin Rittner, Thomas~M Mitchell,
  Adrian~P Jones, Richard Day, John Wheeler, Paul~R Shearing, Pieter Vermeesch,
  and Alberto Striolo.
\newblock Quantifying the anisotropy and tortuosity of permeable pathways in
  clay-rich mudstones using models based on x-ray tomography.
\newblock {\em Scientific reports}, 7(1):1--12, 2017.

\bibitem{ghanbarian2017upscaling}
Behzad Ghanbarian and Farzam Javadpour.
\newblock Upscaling pore pressure-dependent gas permeability in shales.
\newblock {\em Journal of Geophysical Research: Solid Earth},
  122(4):2541--2552, 2017.

\bibitem{woodruff2015measurements}
WF~Woodruff, A~Revil, M~Prasad, and C~Torres-Verd{\'i}n.
\newblock Measurements of elastic and electrical properties of an
  unconventional organic shale under differential loading.
\newblock {\em Geophysics}, 80(4):D363--D383, 2015.

\bibitem{argyris2011structure}
Dimitrios Argyris, Paul~D Ashby, and Alberto Striolo.
\newblock Structure and orientation of interfacial water determine atomic force
  microscopy results: {Insights} from molecular dynamics simulations.
\newblock {\em ACS nano}, 5(3):2215--2223, 2011.

\bibitem{kobayashi2016molecular}
Kazuya Kobayashi, Yunfeng Liang, Ken-ichi Amano, Sumihiko Murata, Toshifumi
  Matsuoka, Satoru Takahashi, Naoya Nishi, and Tetsuo Sakka.
\newblock Molecular dynamics simulation of atomic force microscopy at the
  water--muscovite interface: Hydration layer structure and force analysis.
\newblock {\em Langmuir}, 32(15):3608--3616, 2016.

\bibitem{RN35}
Han Wang, Yuliang Su, Wendong Wang, Guanglong Sheng, Hui Li, and Atif Zafar.
\newblock Enhanced water flow and apparent viscosity model considering
  wettability and shape effects.
\newblock {\em Fuel}, 253:1351--1360, 2019.

\bibitem{RN42}
T~Qiu, XW~Meng, and JP~Huang.
\newblock Nonstraight nanochannels transfer water faster than straight
  nanochannels.
\newblock {\em The Journal of Physical Chemistry B}, 119(4):1496--1502, 2015.

\bibitem{RN43}
Jipeng Li, Xian Kong, Diannan Lu, and Zheng Liu.
\newblock Italicized carbon nanotube facilitating water transport: a molecular
  dynamics simulation.
\newblock {\em Science Bulletin}, 60(18):1580--1586, 2015.

\bibitem{kestin1984thermophysical}
Joseph Kestin, JV~Sengers, B~Kamgar-Parsi, and JMH~Levelt Sengers.
\newblock Thermophysical properties of fluid d2o.
\newblock {\em Journal of Physical and Chemical Reference Data},
  13(2):601--609, 1984.

\bibitem{secchi2016massive}
Eleonora {Secchi}, Sophie {Marbach}, Antoine {Nigu{\`e}s}, Derek {Stein},
  Alessandro {Siria}, and Lyd{\'e}ric {Bocquet}.
\newblock Massive radius-dependent flow slippage in carbon nanotubes.
\newblock {\em Nature}, 537(7619):210--213, 2016.

\bibitem{RN38}
Shuangliang Zhao, Yaofeng Hu, Xiaochen Yu, Yu~Liu, Zhi‐Shan Bai, and Honglai
  Liu.
\newblock Surface wettability effect on fluid transport in nanoscale slit
  pores.
\newblock {\em AIChE Journal}, 63(5):1704--1714, 2017.

\bibitem{RN39}
Fredrik Elwinger, Payam Pourmand, and Istv{\'a}n Fur{\'o}.
\newblock Diffusive transport in pores. tortuosity and molecular interaction
  with the pore wall.
\newblock {\em The Journal of Physical Chemistry C}, 121(25):13757--13764,
  2017.

\bibitem{RN40}
Sen Wang, Qihong Feng, Ming Zha, Farzam Javadpour, and Qinhong Hu.
\newblock Supercritical methane diffusion in shale nanopores: effects of
  pressure, mineral types, and moisture content.
\newblock {\em Energy \& fuels}, 32(1):169--180, 2018.

\bibitem{RN26}
Wei Zhang, Qihong Feng, Sen Wang, and Xiangdong Xing.
\newblock Oil diffusion in shale nanopores: Insight of molecular dynamics
  simulation.
\newblock {\em Journal of Molecular Liquids}, 290:111183, 2019.

\bibitem{RN41}
Choongyeop Lee and Chang-Hwan Choi.
\newblock Structured surfaces for a giant liquid slip.
\newblock {\em Physical Review Letters}, 101(6):064501, 2008.

\bibitem{RN24}
Tuan~A Ho and Yifeng Wang.
\newblock Enhancement of oil flow in shale nanopores by manipulating friction
  and viscosity.
\newblock {\em Physical Chemistry Chemical Physics}, 21(24):12777--12786, 2019.

\bibitem{granick2003slippery}
Steve Granick, Yingxi Zhu, and Hyunjung Lee.
\newblock Slippery questions about complex fluids flowing past solids.
\newblock {\em Nature materials}, 2(4):221--227, 2003.

\bibitem{carman1956flow}
Philip~Crosbie Carman.
\newblock {\em Flow of gases through porous media}.
\newblock Butterworths Scientific Publications, London, 1956.

\bibitem{amaefule1993enhanced}
Jude~O Amaefule, Mehmet Altunbay, Djebbar Tiab, David~G Kersey, Dare~K Keelan,
  et~al.
\newblock Enhanced reservoir description: using core and log data to identify
  hydraulic (flow) units and predict permeability in uncored intervals/wells.
\newblock In {\em SPE annual technical conference and exhibition}. Society of
  Petroleum Engineers, 1993.

\bibitem{carman1937fluid}
Philip~Crosbie Carman.
\newblock Fluid flow through granular beds.
\newblock {\em Trans. Inst. Chem. Eng.}, 15:150--166, 1937.

\bibitem{darabi2012gas}
Hamed Darabi, Amin Ettehadtavakkol, F.~Javadpour, and K.~Sepehrnoori.
\newblock Gas flow in ultra-tight shale strata.
\newblock {\em Journal of Fluid Mechanics}, 710:641--658, 2012.

\bibitem{fan2020accurate}
Dian Fan, Anh Phan, and Alberto Striolo.
\newblock Accurate permeability prediction in tight gas rocks via lattice
  boltzmann simulations with an improved boundary condition.
\newblock {\em Journal of Natural Gas Science and Engineering}, 73:103049,
  2020.

\bibitem{javadpour2015gas}
Farzam Javadpour and Amin Ettehadtavakkol.
\newblock Gas transport processes in shale.
\newblock {\em Fundamentals of gas shale reservoirs}, pages 245--266, 2015.

\bibitem{fan2017semi}
Dian Fan and Amin Ettehadtavakkol.
\newblock Semi-analytical modeling of shale gas flow through fractal induced
  fracture networks with microseismic data.
\newblock {\em Fuel}, 193:444--459, 2017.

\bibitem{zisman1964relation}
W.~A. Zisman.
\newblock {\em Relation of the Equilibrium Contact Angle to Liquid and Solid
  Constitution}, chapter~1, pages 1--51.
\newblock ACS Publications, 1964.

\bibitem{klinkenberg1941permeability}
LJ~Klinkenberg et~al.
\newblock The permeability of porous media to liquids and gases.
\newblock In {\em Drilling and production practice}. American Petroleum
  Institute, 1941.

\bibitem{wu2017apparent}
Lei Wu, Minh~Tuan Ho, Lefki Germanou, Xiao-Jun Gu, Chang Liu, Kun Xu, and
  Yonghao Zhang.
\newblock On the apparent permeability of porous media in rarefied gas flows.
\newblock {\em Journal of Fluid Mechanics}, 822:398--417, 2017.

\bibitem{arkilic2001mass}
Errol~B Arkilic, Kenneth~S Breuer, and Martin~A Schmidt.
\newblock Mass flow and tangential momentum accommodation in silicon
  micromachined channels.
\newblock {\em Journal of fluid mechanics}, 437:29--43, 2001.

\bibitem{beskok1999report}
Ali Beskok and George~Em Karniadakis.
\newblock Report: a model for flows in channels, pipes, and ducts at micro and
  nano scales.
\newblock {\em Microscale Thermophysical Engineering}, 3(1):43--77, 1999.

\bibitem{yu2002fractal}
Boming Yu and Ping Cheng.
\newblock A fractal permeability model for bi-dispersed porous media.
\newblock {\em International Journal of Heat and Mass Transfer},
  45(14):2983--2993, 2002.

\bibitem{RN25}
Shu Yang, Hassan Dehghanpour, Mojtaba Binazadeh, and Pingchuan Dong.
\newblock A molecular dynamics explanation for fast imbibition of oil in
  organic tight rocks.
\newblock {\em Fuel}, 190:409--419, 2017.

\bibitem{RN27}
Hari~Krishna Chilukoti, Gota Kikugawa, and Taku Ohara.
\newblock Mass transport and structure of liquid n-alkane mixtures in the
  vicinity of alpha-quartz substrates.
\newblock {\em RSC advances}, 6(102):99704--99713, 2016.

\bibitem{RN28}
Tue Hassenkam, Lone~Lindbæk Skovbjerg, and Susan Louise~Svane Stipp.
\newblock Probing the intrinsically oil-wet surfaces of pores in north sea
  chalk at subpore resolution.
\newblock {\em Proceedings of the National Academy of Sciences},
  106(15):6071--6076, 2009.

\bibitem{RN29}
Yinan Hu, Deepak Devegowda, Alberto Striolo, Anh Phan, Tuan~A Ho, Faruk Civan,
  and Richard~F Sigal.
\newblock Microscopic dynamics of water and hydrocarbon in shale-kerogen pores
  of potentially mixed wettability.
\newblock {\em Spe Journal}, 20(01):112--124, 2014.

\end{thebibliography}

\end{document}